\documentclass[twocolumn,superscriptaddress,showpacs,preprintnumbers,amsmath,amssymb,nofootinbib,prd]{revtex4-1}

\usepackage{graphicx}
\usepackage{dcolumn}
\usepackage{bm}
\usepackage[dvips]{color}
\usepackage{epsfig}

\RequirePackage{xspace}

\newcommand{\dsdpi}{\ensuremath{D^{*\pm}\to D\pi^{\pm}}\xspace}

\newcommand{\kspp}{\ensuremath{K^0_S\pi^+\pi^-}\xspace}
\newcommand{\dkpp}{\ensuremath{D\to K^0_S\pi^+\pi^-}\xspace}
\newcommand{\dnkpp}{\ensuremath{D^0\to K^0_S\pi^+\pi^-}\xspace}
\newcommand{\dbkpp}{\ensuremath{\overline{D}{}^0\to K^0_S\pi^+\pi^-}\xspace}

\newcommand{\bdk}{\ensuremath{B^{\pm}\to D K^{\pm}}\xspace}

\newcommand{\bdkm}{\ensuremath{B^-\to D K^-}\xspace}
\newcommand{\bdkp}{\ensuremath{B^+\to D K^+}\xspace}

\newcommand{\bdpi}{\ensuremath{B^{\pm}\to D \pi^{\pm}}\xspace}

\newcommand{\bdpim}{\ensuremath{B^-\to D \pi^-}\xspace}
\newcommand{\bdpip}{\ensuremath{B^+\to D \pi^+}\xspace}

\newcommand{\dn}{\ensuremath{D^0}\xspace}
\newcommand{\dnbar}{\ensuremath{\overline{D}{}^0}\xspace}

\newcommand{\ab}{\ensuremath{A_B}\xspace}

\newcommand{\pd}{\ensuremath{P}\xspace}
\newcommand{\pdbar}{\ensuremath{\overline{P}}\xspace}

\newcommand{\ad}{\ensuremath{A}\xspace}
\newcommand{\adbar}{\ensuremath{\overline{A}}\xspace}
\newcommand{\dvar}{\ensuremath{(m^2_{+}, m^2_{-})}\xspace}

\newcommand{\aad}{\ensuremath{|A|}\xspace}
\newcommand{\aadbar}{\ensuremath{|\overline{A}|}\xspace}

\newcommand{\mbc}{\ensuremath{M_{\rm bc}}\xspace}
\newcommand{\de}{\ensuremath{\Delta E}\xspace}
\newcommand{\thr}{\ensuremath{\cos\theta_{\rm thr}}\xspace}
\newcommand{\fish}{\ensuremath{\mathcal{F}}\xspace}

\renewcommand{\deg}{\ensuremath{^{\circ}}\xspace}

\newcommand{\ddd}{\ensuremath{\Delta\delta_D}\xspace}

\begin{document}

\preprint{Belle Preprint 2012-12}
\preprint{KEK Preprint 2012-4}

\title{First Measurement of \boldmath{$\phi_3$} with a 
Model-independent Dalitz Plot Analysis 
of \boldmath{\bdk}, \boldmath{\dkpp} Decay}

\affiliation{University of Bonn, Bonn}
\affiliation{Budker Institute of Nuclear Physics SB RAS and Novosibirsk State University, Novosibirsk 630090}
\affiliation{Faculty of Mathematics and Physics, Charles University, Prague}
\affiliation{University of Cincinnati, Cincinnati, Ohio 45221}
\affiliation{Department of Physics, Fu Jen Catholic University, Taipei}
\affiliation{Hanyang University, Seoul}
\affiliation{University of Hawaii, Honolulu, Hawaii 96822}
\affiliation{High Energy Accelerator Research Organization (KEK), Tsukuba}
\affiliation{Hiroshima Institute of Technology, Hiroshima}
\affiliation{Indian Institute of Technology Guwahati, Guwahati}
\affiliation{Indian Institute of Technology Madras, Madras}
\affiliation{Institute of High Energy Physics, Chinese Academy of Sciences, Beijing}
\affiliation{Institute of High Energy Physics, Vienna}
\affiliation{Institute of High Energy Physics, Protvino}
\affiliation{Institute for Theoretical and Experimental Physics, Moscow}
\affiliation{J. Stefan Institute, Ljubljana}
\affiliation{Kanagawa University, Yokohama}
\affiliation{Institut f\"ur Experimentelle Kernphysik, Karlsruher Institut f\"ur Technologie, Karlsruhe}
\affiliation{Korea Institute of Science and Technology Information, Daejeon}
\affiliation{Korea University, Seoul}
\affiliation{Kyungpook National University, Taegu}
\affiliation{\'Ecole Polytechnique F\'ed\'erale de Lausanne (EPFL), Lausanne}
\affiliation{Faculty of Mathematics and Physics, University of Ljubljana, Ljubljana}
\affiliation{Luther College, Decorah, Iowa 52101}
\affiliation{University of Maribor, Maribor}
\affiliation{Max-Planck-Institut f\"ur Physik, M\"unchen}
\affiliation{University of Melbourne, School of Physics, Victoria 3010}
\affiliation{Graduate School of Science, Nagoya University, Nagoya}
\affiliation{Kobayashi-Maskawa Institute, Nagoya University, Nagoya}
\affiliation{Nara Women's University, Nara}
\affiliation{National United University, Miao Li}
\affiliation{Department of Physics, National Taiwan University, Taipei}
\affiliation{H. Niewodniczanski Institute of Nuclear Physics, Krakow}
\affiliation{Nippon Dental University, Niigata}
\affiliation{Niigata University, Niigata}
\affiliation{University of Nova Gorica, Nova Gorica}
\affiliation{Osaka City University, Osaka}
\affiliation{Pacific Northwest National Laboratory, Richland, Washington 99352}
\affiliation{Panjab University, Chandigarh}
\affiliation{Research Center for Nuclear Physics, Osaka University, Osaka}
\affiliation{University of Science and Technology of China, Hefei}
\affiliation{Seoul National University, Seoul}
\affiliation{Sungkyunkwan University, Suwon}
\affiliation{School of Physics, University of Sydney, NSW 2006}
\affiliation{Tata Institute of Fundamental Research, Mumbai}
\affiliation{Excellence Cluster Universe, Technische Universit\"at M\"unchen, Garching}
\affiliation{Toho University, Funabashi}
\affiliation{Tohoku Gakuin University, Tagajo}
\affiliation{Tohoku University, Sendai}
\affiliation{Department of Physics, University of Tokyo, Tokyo}
\affiliation{Tokyo Institute of Technology, Tokyo}
\affiliation{Tokyo Metropolitan University, Tokyo}
\affiliation{Tokyo University of Agriculture and Technology, Tokyo}
\affiliation{CNP, Virginia Polytechnic Institute and State University, Blacksburg, Virginia 24061}
\affiliation{Wayne State University, Detroit, Michigan 48202}
\affiliation{Yamagata University, Yamagata}
\affiliation{Yonsei University, Seoul}
  \author{H.~Aihara}\affiliation{Department of Physics, University of Tokyo, Tokyo} 
  \author{K.~Arinstein}\affiliation{Budker Institute of Nuclear Physics SB RAS and Novosibirsk State University, Novosibirsk 630090} 
  \author{D.~M.~Asner}\affiliation{Pacific Northwest National Laboratory, Richland, Washington 99352} 
  \author{V.~Aulchenko}\affiliation{Budker Institute of Nuclear Physics SB RAS and Novosibirsk State University, Novosibirsk 630090} 
  \author{T.~Aushev}\affiliation{Institute for Theoretical and Experimental Physics, Moscow} 
  \author{A.~M.~Bakich}\affiliation{School of Physics, University of Sydney, NSW 2006} 
  \author{K.~Belous}\affiliation{Institute of High Energy Physics, Protvino} 
  \author{B.~Bhuyan}\affiliation{Indian Institute of Technology Guwahati, Guwahati} 
  \author{M.~Bischofberger}\affiliation{Nara Women's University, Nara} 
  \author{A.~Bondar}\affiliation{Budker Institute of Nuclear Physics SB RAS and Novosibirsk State University, Novosibirsk 630090} 
% ...
\collaboration{The Belle Collaboration}


\begin{abstract} 

We present the first measurement of the angle $\phi_3$ of the 
Unitarity Triangle using a model-independent Dalitz plot analysis 
of \bdk, \dkpp decays. The method uses, as input, measurements of 
the strong phase of the \dkpp amplitude from the CLEO collaboration. 
The result is based on the full data set of 
$772\times 10^6$ $B\overline{B}$ pairs collected by the Belle 
experiment at the $\Upsilon(4S)$ resonance. 
We obtain $\phi_3 = (77.3^{+15.1}_{-14.9} \pm 4.1 \pm 4.3)^{\circ}$
and the suppressed amplitude ratio 
$r_B = 0.145\pm 0.030 \pm 0.010\pm 0.011$. Here the first error 
is statistical, the second is the experimental systematic uncertainty, 
and the third is the error due to the precision of the strong-phase 
parameters obtained by CLEO. 
\end{abstract}
\maketitle

\section{Introduction}

The angle $\phi_3$ (also denoted as $\gamma$) is one of the least 
well-constrained parameters of the Unitarity Triangle. 
The measurement that currently dominates sensitivity to $\phi_3$ uses 
$\bdk$ decays with the neutral $D$ meson decaying to a three-body final 
state such as $\kspp$~\cite{giri, binp_dalitz}. 
The weak phase $\phi_3$ appears in the 
interference between $b\to c\bar{u}s$ and $b\to u\bar{c}s$ transitions. 
The value of $\phi_3$ is determined by exploiting differences between the 
\kspp Dalitz plots for $D$ mesons from $B^{+}$ and $B^{-}$ decay. 
Theoretical uncertainties in the $\phi_3$ determination in \bdk decays 
are expected to be negligible~\cite{zupan}, and the main difficulty in its measurement 
is the very low probability of the decays that are involved. However, the method 
based on Dalitz plot analysis requires the knowledge of the amplitude of the 
\dnkpp decay, including its complex phase. The amplitude can be obtained from a 
model that involves isobar and K-matrix~\cite{kmatrix} descriptions of 
the decay dynamics, and thus results in a model uncertainty for the $\phi_3$
measurement. In the latest model-dependent Dalitz plot analyses 
performed by BaBar and Belle, this uncertainty ranges from 3$^{\circ}$ 
to 9$^{\circ}$~\cite{babar_gamma_1, babar_gamma_2, babar_gamma_3, belle_phi3_1, belle_phi3_2, belle_phi3_3}. 

A method to eliminate the model uncertainty using a 
binned Dalitz plot analysis has been proposed
by Giri {\it et al.}~\cite{giri}. Information about the strong 
phase in the \dnkpp decay is obtained from the decays of quantum-correlated
$D^0$ pairs produced in the $\psi(3770)\to D^0\overline{D}{}^0$ process. 
As a result, the model uncertainty is replaced by a statistical 
error related to the precision of the strong-phase parameters. 
This method has been further developed in Refs.~\cite{modind2006, modind2008}, 
where its experimental feasibility has been shown 
along with a proposed analysis procedure to optimally 
use the available $B$ decays and correlated $D^0$ pairs. 
In this paper, we report the first measurement of $\phi_3$ 
using a model-independent Dalitz plot analysis of the \dkpp 
decay from the mode \bdk, based on a 711 fb$^{-1}$ data sample
(corresponding to $772\times 10^6$ $B\overline{B}$ pairs)
collected by the Belle detector at the KEKB asymmetric-energy
$e^+e^-$ collider. 
This analysis uses the recent measurement of the strong phase in 
\dnkpp and $D^0\to K^0_SK^+K^-$ decays performed by the CLEO 
collaboration~\cite{cleo_1, cleo_2}.

\section{The model-independent Dalitz plot analysis technique}

The amplitude of the \bdkp, \dkpp decay is a superposition
of the $B^+\to \overline{D}{}^0K^+$ and $B^+\to D^0K^+$ amplitudes
\begin{equation}
  \ab\dvar=\adbar+ r_Be^{i(\delta_B+\phi_3)}\ad\,, 
\end{equation}
where $m^2_+$ and $m^2_-$ are the Dalitz plot variables --- the 
squared invariant masses of $K_S^0\pi^+$ and $K_S^0\pi^-$ combinations, 
respectively,
$\adbar=\adbar\dvar$ is the amplitude of the \dbkpp decay,  $\ad=\ad\dvar$ 
is the amplitude of the \dnkpp decay, 
$r_B$ is the ratio of the absolute values of 
the $B^+\to \overline{D}{}^0K^+$ and $B^+\to D^0K^+$
amplitudes, and $\delta_B$ is the strong-phase difference between 
them. In the case of $CP$ conservation in the $D$ decay 
$\ad\dvar = \adbar(m^2_{-},m^2_{+})$. 
The Dalitz plot density of the $D$ decay from \bdkp\ is given by 
\begin{equation}
 \begin{split}
  P_{B}=|\ab|^2 = & |\adbar+ r_Be^{i(\delta_B+\phi_3)}\ad|^2 = \\
  & \pdbar +r_B^2\pd + 2\sqrt{\pd\pdbar}(x_+ C+y_+ S)\,, 
 \label{p_b}
 \end{split}
\end{equation}
where $\pd\dvar=|\ad|^2$, $\pdbar\dvar=|\adbar|^2$; while
\begin{equation}
  x_+ = r_B\cos(\delta_B+\phi_3)\,, \;\;\;
  y_+ = r_B\sin(\delta_B+\phi_3)\,;
\end{equation}
and the functions $C=C\dvar$ and $S=S\dvar$ are the cosine and sine of the 
strong-phase difference $\delta_D\dvar=\arg\adbar-\arg\ad$ between the \dbkpp and \dnkpp 
amplitudes.\footnote{
  This paper follows the convention for strong phases in $D$ decay
  amplitudes introduced in Ref.~\cite{modind2008}. 
}
The equations for the charge-conjugate mode \bdkm are obtained 
with the substitution $\phi_3 \rightarrow -\phi_3$ and 
$A\leftrightarrow \overline{A}$; the corresponding parameters 
that depend on the $B^-$ decay amplitude are: 
\begin{equation}
  x_- = r_B\cos(\delta_B-\phi_3)\,, \;\;\;
  y_- = r_B\sin(\delta_B-\phi_3)\,.
\end{equation}
Using both $B$ charges, one can obtain $\phi_3$ and $\delta_B$ separately. 

Up to this point, the description of the model-dependent and 
model-independent techniques is the same. The model-dependent analysis deals
directly with the Dalitz plot density, and the functions $C$ and $S$ 
are obtained from model assumptions in the fit to the \dnkpp amplitude. 
In the model-independent approach, the Dalitz plot is divided into 
$2\mathcal{N}$ bins symmetric under the exchange $m^2_-\leftrightarrow m^2_+$. 
The bin index $i$ ranges from $-\mathcal{N}$ to $\mathcal{N}$ (excluding 0); 
the exchange $m^2_+ \leftrightarrow m^2_-$ corresponds to the exchange 
$i\leftrightarrow -i$. The expected number of events in bin $i$ 
of the Dalitz plot of the $D$ meson from \bdk is
\begin{equation}
  N^{\pm}_i = 
  h_{B} \left[
    K_{\pm i} + r_B^2K_{\mp i} + 2\sqrt{K_iK_{-i}}(x_{\pm} c_i \pm y_{\pm} s_i)
  \right] \,, 
  \label{n_b}
\end{equation}
where $h_{B}$ is a normalization constant and $K_i$ is the number of events 
in the $i^{\mathrm{th}}$ bin of the \kspp Dalitz plot of the $D$ meson 
in a flavor eigenstate. A sample of flavor-tagged $D^0$ mesons is 
obtained by reconstructing $D^{*\pm}\to D\pi^\pm$ decays (note that 
charge conjugation is assumed throughout this paper unless otherwise 
stated). The terms $c_i$ and $s_i$ include information about 
the functions $C$ and $S$ averaged over the bin region:
\begin{equation}
  c_i=\frac{\int\limits_{\mathcal{D}_i}
            \aad\aadbar
            C\,d\mathcal{D}
            }{\sqrt{
            \int\limits_{\mathcal{D}_i}\aad^2 d\mathcal{D}
            \int\limits_{\mathcal{D}_i}\aadbar^2 d\mathcal{D}
            }}\,. 
  \label{cs}
\end{equation}
Here $\mathcal{D}$ represents the Dalitz plot phase space and 
$\mathcal{D}_i$ is the bin region over which the integration is performed. 
The terms $s_i$ are defined similarly with $C$ substituted by $S$. 
The absence of $CP$ violation in the $D$ decay implies $c_i = c_{-i}$ and
$s_i = - s_{-i}$. 

The values of the $c_i$ and $s_i$ terms are measured 
in the quantum correlations of $D$ pairs by charm-factory experiments 
operating at the threshold of $D\overline{D}$ pair 
production~\cite{cleo_1, cleo_2}. The measurement involves 
studies of the four-dimensional (4D) density of two correlated \dkpp Dalitz plots, 
as well as decays of a $D$ meson tagged in a $CP$-eigenstate decaying to $K^0_S\pi^+\pi^-$. 
The wave function of the two mesons is antisymmetric, thus 
the 4D density of two correlated \dkpp Dalitz plots is
\begin{equation}
\begin{split}
  |A_{\rm corr}&(m_+^2,m_-^2,m'^2_+,m'^2_-)|^2=|\ad\adbar{}'-\adbar\ad'|^2=\\
     &\pd\pdbar{}' + \pdbar \pd' -
     2\sqrt{\pd\pdbar{}'\pdbar\pd'}(C C'+S S')\,, 
  \label{p_corr}
\end{split}
\end{equation}
where the primed and unprimed quantities correspond to the two decaying $D$ mesons. 
Similarly, the density of the decay \dkpp, where the $D$ meson is in a $CP$-eigenstate, is
\begin{equation}
\begin{split}
  |A_{\rm CP}(m_+^2,m_-^2)|^2=|\ad\pm\adbar|^2=
     \pd + \pdbar \pm
     2\sqrt{\pd\pdbar}C\,. 
  \label{p_cp}
\end{split}
\end{equation}
CLEO uses these relations to obtain $c_i$ and $s_i$ values. Once they 
are measured, the system of equations (\ref{n_b})
contains only three free parameters ($x$, $y$, and $h_B$)
for each $B$ charge, and can be solved using a maximum likelihood 
method to extract the value of $\phi_3$. 

We have neglected charm-mixing effects in $D$ decays from 
both the \bdk process and in quantum-correlated $D\overline{D}$ production. 
It has been shown~\cite{modind_mixing} that although the charm 
mixing correction is of first order in the mixing parameters $x_D, y_D$, 
it is numerically small (of the order $0.2^{\circ}$ for $x_D,y_D\sim 0.01$) 
and can be neglected at the current level of precision. 
Future precision measurements of $\phi_3$
can account for charm mixing and $CP$ violation (both in mixing 
and decay) using the measurement of the corresponding parameters. 

In principle, the set of relations defined by Eq.~(\ref{n_b}) 
can be solved without external constraints on $c_i$ and $s_i$ 
for $\mathcal{N} \ge 2$.
However, due to the small value of $r_B$, there is very little
sensitivity to the $c_i$ and $s_i$ parameters in \bdk decays, which 
results in a reduction in the precision on $\phi_3$ that can be 
obtained~\cite{modind2006}.

\section{CLEO input}

The procedure for a binned Dalitz plot analysis should give the correct 
results for any binning. However, the statistical accuracy depends strongly 
on the amplitude behavior across the bins. Large variations of 
the amplitude within a bin result in loss of coherence in the interference term. 
This effect becomes especially significant with limited statistics when a 
small number of bins must be used to ensure a stable fit. Greater statistical 
precision is obtained for the binning in which the phase difference 
between the \dn and \dnbar amplitudes varies 
as little as possible within a bin~\cite{modind2008}. For optimal precision, 
one also has to take the variations of the absolute value of the amplitude 
into account, along with contributions from background events. 
The procedure to optimize the binning for the maximal  
statistical precision of $\phi_3$ has been proposed in Ref.~\cite{modind2008} and 
generalized to the case with background in Ref.~\cite{cleo_2}. 
It has been shown that as few as 16 bins are enough 
to reach a statistical precision that is only 10--20\% worse than in the 
unbinned case. 

\begin{figure}
  \begin{center}
  \includegraphics[width=0.37\textwidth]{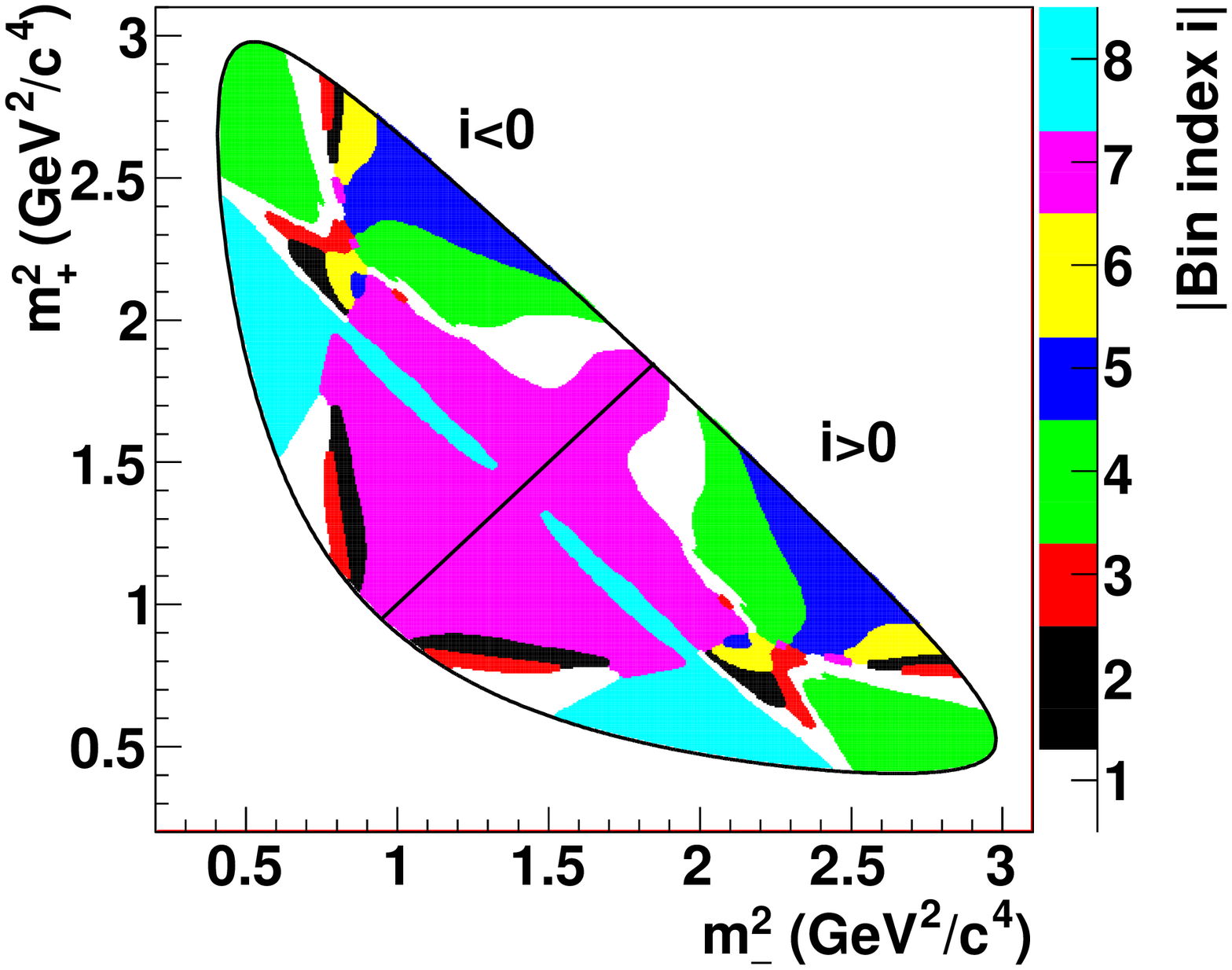}
  \end{center}
  \vspace{-\baselineskip}
  \caption{Optimal binning of the \dkpp Dalitz plot. 
    The color scale indicated corresponds to the 
    absolute value of the bin index, $|i|$.}
  \label{fig:babar_opt_bins}
\end{figure}

The optimization of binning sensitivity uses the amplitude of the \dkpp 
decay. It should be noted, however, that although the choice of 
binning is model-dependent, a poor choice of model results only in a loss 
of precision, not bias, of the measured parameters~\cite{modind2008}. 
CLEO measured $c_i$ and $s_i$ parameters for four different binnings 
with $\mathcal{N}=8$: 
\begin{enumerate}
  \item Bins equally distributed in the phase 
        difference $\Delta\delta_D$ between the \dn and \dnbar 
        decay amplitudes, with the amplitude from the BaBar 
        measurement~\cite{babar_gamma_2}. 
  \item Same as option 1, but with the amplitude 
        from the Belle analysis~\cite{belle_phi3_3}. 
  \item Optimized for statistical precision according 
        to the procedure from~\cite{modind2008} 
        (see Fig.~\ref{fig:babar_opt_bins}). The effect of the 
        background in $B$ data is not taken into account in the optimization. 
        The amplitude is taken from the BaBar measurement~\cite{babar_gamma_2}. 
  \item Same as option 3, but optimized for an analysis with high 
        background in $B$ data ({\it e.~g.,} at LHCb). 
\end{enumerate}

Our analysis uses the optimal binning shown in Fig.~\ref{fig:babar_opt_bins}
(option 3) as the baseline since it offers better statistical accuracy. 
In addition, we use the equal phase difference binning (\ddd-binning, 
option 1) as a cross-check. 

\begin{figure}
  \begin{center}
  \includegraphics[width=0.37\textwidth]{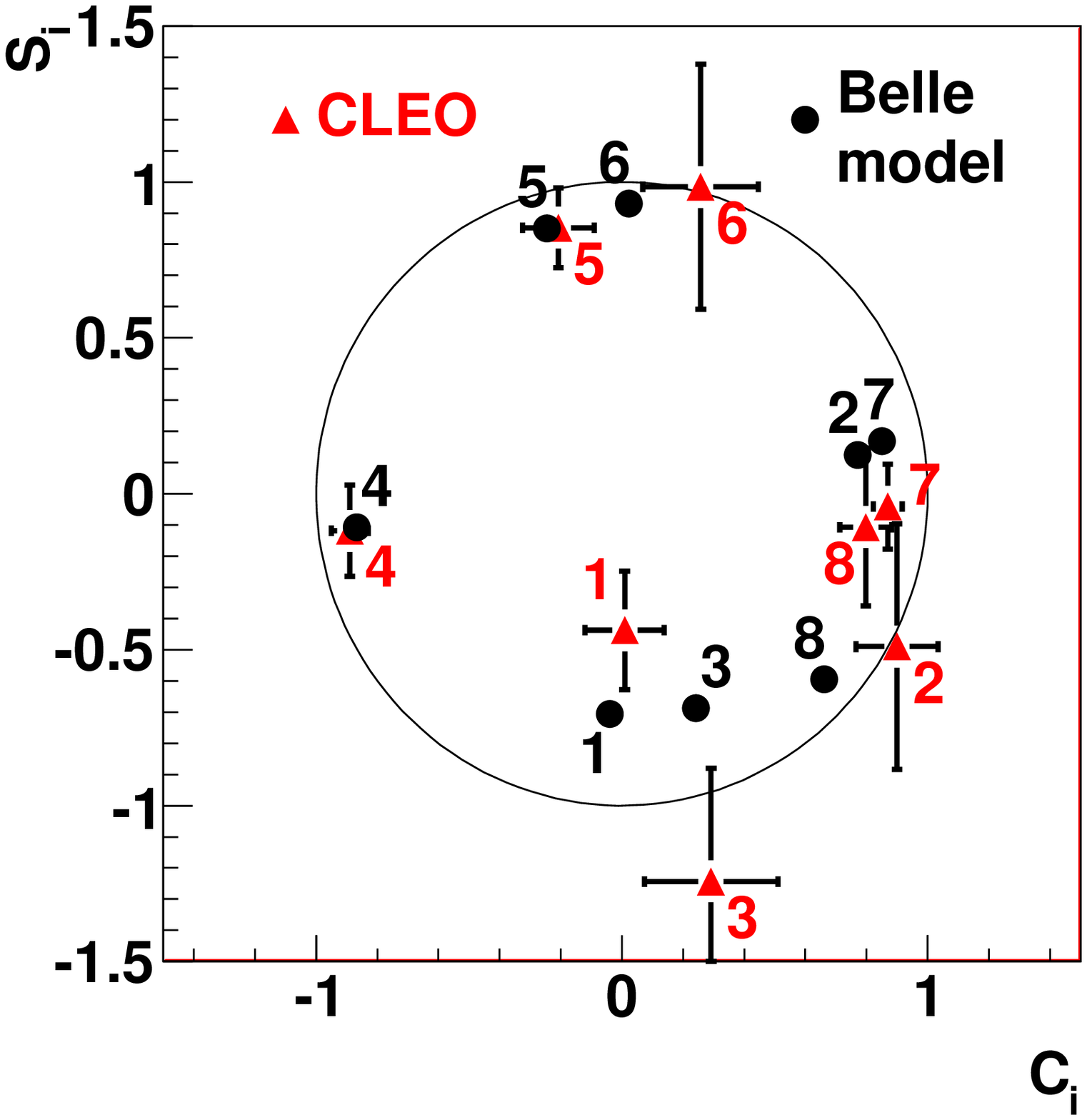}
  \end{center}
  \vspace{-\baselineskip}
  \caption{Comparison of phase terms $(c_i, s_i)$ for the optimal binning
           measured by CLEO, and calculated from the Belle \dkpp amplitude model. }
  \label{fig:cs_cleo_opt}
\end{figure}

The results of the CLEO measurement of $c_i$ and $s_i$ for the optimal 
binning are presented in Table~\ref{tab:cs_cleo_opt}. The same results 
in graphical form are shown in Fig.~\ref{fig:cs_cleo_opt}. 
The values of $c_i$ and $s_i$ calculated from the Belle model~\cite{belle_phi3_3}
are compared to the measurements and are found to be in reasonable 
agreement with $\chi^2=18.6$ for the number of degrees of freedom $\mbox{ndf}=16$
(the corresponding $p$-value is $p=29\%$).

\begin{table}
  \caption{Values of $c_i$ and $s_i$ for the optimal 
           binning measured by CLEO~\cite{cleo_2}, 
           and calculated from the Belle \dkpp amplitude model. }
  \label{tab:cs_cleo_opt}
  \begin{tabular}{|l|c|c|}
    \hline
                 &           CLEO measurement & Belle model    \\
    \hline
    $c_1$        & $-0.009\pm 0.088\pm 0.094$ & $-0.039$         \\
    $c_2$        & $+0.900\pm 0.106\pm 0.082$ & $+0.771$         \\
    $c_3$        & $+0.292\pm 0.168\pm 0.139$ & $+0.242$         \\
    $c_4$        & $-0.890\pm 0.041\pm 0.044$ & $-0.867$         \\
    $c_5$        & $-0.208\pm 0.085\pm 0.080$ & $-0.246$         \\
    $c_6$        & $+0.258\pm 0.155\pm 0.108$ & $+0.023$         \\
    $c_7$        & $+0.869\pm 0.034\pm 0.033$ & $+0.851$         \\
    $c_8$        & $+0.798\pm 0.070\pm 0.047$ & $+0.662$         \\
    \hline
    $s_1$        & $-0.438\pm 0.184\pm 0.045$ & $-0.706$         \\
    $s_2$        & $-0.490\pm 0.295\pm 0.261$ & $+0.124$         \\
    $s_3$        & $-1.243\pm 0.341\pm 0.123$ & $-0.687$         \\
    $s_4$        & $-0.119\pm 0.141\pm 0.038$ & $-0.108$         \\
    $s_5$        & $+0.853\pm 0.123\pm 0.035$ & $+0.851$         \\
    $s_6$        & $+0.984\pm 0.357\pm 0.165$ & $+0.930$         \\
    $s_7$        & $-0.041\pm 0.132\pm 0.034$ & $+0.169$         \\
    $s_8$        & $-0.107\pm 0.240\pm 0.080$ & $-0.596$         \\
    \hline
  \end{tabular}
\end{table}

As is apparent from Fig.~\ref{fig:cs_cleo_opt}, the chosen binning contains 
bins where the strong phase difference between $D^0$ and $\overline{D}{}^0$ amplitudes 
is close to zero (bins with $|i|=2,7,8$) and $180^{\circ}$ (bin with $|i|=4$) which provide 
sensitivity to $x_{\pm}$, as well as bins with the strong phase difference close 
to $90^{\circ}$ and $270^{\circ}$ (bins with $|i|=1,3,5,6$), more sensitive to $y_{\pm}$. 
This ensures that the method is sensitive to $\phi_3$ for any combination of $\phi_3$
and $\delta_B$ values. 

\section{Analysis procedure}

Equation (\ref{n_b}) is the key relation used in the analysis, but it only 
holds if there is no background, a uniform Dalitz plot acceptance and 
no cross-feed between bins. (Cross-feed is due to invariant-mass 
resolution and radiative corrections.) 
In this section we outline the procedures that account for these 
experimental effects.

\subsection{Efficiency profile}

\label{sec:eff}

We note that the Eqs.~(\ref{p_b}), (\ref{p_corr}) and (\ref{p_cp}) do not change 
after the transformation $P\to \epsilon P$ when the efficiency profile 
$\epsilon(m^2_+, m^2_-)$ is symmetric: 
$\epsilon(m^2_+, m^2_-)=\epsilon(m^2_-, m^2_+)$. This implies that if the 
efficiency profile is the same in all of the three modes involved in the measurement
(flavor $D$, correlated $\psi(3770)\to D\overline{D}$, and $D$ from $B\to DK$), 
the result will be unbiased even if no efficiency correction 
is applied. 

The effect of non-uniform efficiency over the Dalitz plot cancels out 
when using a flavor-tagged $D$ sample with kinematic properties that are
similar to the sample for the signal $B$ decay. 
This approach allows for the removal of 
systematic error associated with the possible inaccuracy in the description 
of the detector acceptance in the Monte Carlo (MC) simulation. 
The center-of-mass (CM) $D$ momentum 
distribution for $B\to DK$ decays is practically uniform in the narrow 
range $2.10\mbox{ GeV}/c < p_D < 2.45\mbox{ GeV}/c$. We assume that 
the efficiency profile depends mostly on the $D$ momentum and take the 
flavor-tagged sample with an average momentum of $p_D=2.3$ GeV/$c$
(we use a wider range of $D$ momenta than in $B\to DK$ to increase the 
statistics). The assumption that the efficiency profile depends only 
on the $D$ momentum is tested using MC simulation, and the remaining 
difference is treated as a systematic uncertainty. 

While calculating $c_i$ and $s_i$, CLEO applies an efficiency correction, 
therefore the values reported in their analysis correspond to a flat 
efficiency profile. 
To use the $c_i$ and $s_i$ values in the $\phi_3$ analysis, they have to be 
corrected for the Belle efficiency profile. This correction cannot be 
performed in a completely model-independent way, since the correction terms 
include the phase variation inside the bin. Fortunately, the calculations
using the Belle \dkpp model show that this correction is negligible even 
for very large non-uniformity of the efficiency profile. The difference 
between the uncorrected $c_i$ and $s_i$ terms and those corrected for the 
efficiency, calculated using the efficiency profile parameterization 
used in the 605 fb$^{-1}$ analysis~\cite{belle_phi3_3}, 
does not exceed 0.01, which is negligible compared to the 
statistical error.  

\subsection{Momentum resolution}

Momentum resolution leads to migration of events between the bins. 
In the binned approach, this effect can be corrected in a 
non-parametric way. The migration can be described by a linear transformation 
of the number of events in bins:
\begin{equation}
N'_i = \sum \alpha_{ik} N_k, 
\end{equation}
where $N_k$ is the number of events that bin $k$ would contain without the 
cross-feed, and $N'_i$ is the reconstructed number of events in bin $i$. 
The cross-feed matrix $\alpha_{ik}$ is nearly a unit matrix: $\alpha_{ik} \ll 1$ for 
$i\neq k$. It is obtained from a signal MC simulation generated with the amplitude
model reported in Ref~\cite{belle_phi3_3}. In the case 
of a \dkpp decay from a $B$, the cross-feed depends on the parameters $x$ and $y$. 
However, this is a minor correction 
to an already small effect due to cross-feed; therefore it is neglected.

Migration of events between the bins also occurs due to final state 
radiation (FSR). The $c_i$ and $s_i$ terms in the CLEO measurement are not 
corrected for FSR; we therefore do not simulate FSR to obtain the 
cross-feed matrix to minimize the bias due to this effect. Comparison of 
the cross-feed with and without FSR shows that this effect is negligible. 

\subsection{Fit procedure}

\label{sec:fit_procedure}

The background contribution has to be accounted for in the calculation 
of the values $N_i$ and $K_i$. Statistically the most effective way of 
calculating the number of signal events (especially in the case of $N_i$, 
where the statistics is a limiting factor) is to perform,
in each bin $i$ of the Dalitz plot, an unbinned fit 
in the variables used to distinguish the signal from the background. 

Two different approaches are used in this analysis to obtain the $CP$ violating 
parameters from the data: separate fits in bins, and a combined fit.

In the first, 
we fit the data distribution in each bin separately, with the 
number of events for signal and backgrounds as free parameters. 
Once the numbers of events in bins $N_i$ are found, we 
use them in Eq.~\ref{n_b} to obtain the parameters $(x_{\pm}, y_{\pm})$.
This is accomplished by minimizing a negative logarithmic likelihood of the form
\begin{equation}
          -2\log\mathcal{L}(x,y)=
             -2\sum_i\log p(\langle N_i\rangle(x,y),N_i,\sigma_{N_i}),
\end{equation}
where $\langle N_i\rangle(x,y)$ is the expected number of events in 
the bin $i$ obtained from Eq.~\ref{n_b}. Here, $N_i$ and $\sigma_{N_i}$
are the observed number of events in data and 
the uncertainty on $N_i$, respectively.
If the probability density function (PDF) $p$ is Gaussian, 
this procedure is equivalent to a $\chi^2$ fit; however, the assumption 
of the Gaussian distribution may introduce a bias in the case of
low statistics in certain bins. 

The procedure described above does not make any assumptions on the 
Dalitz distribution of the background events, since the fits in each bin are
independent. Thus there is no associated systematic uncertainty. However, 
in the case of a small number of events and many background components this 
can be a limiting factor. Our second approach is to use the combined 
fit with a common likelihood for all bins. The relative numbers of background
events in bins in such a fit can be constrained externally from
MC and control samples. In addition, for the case of the combined fit, the 
two-step procedure of first extracting the numbers of signal events, and 
then using them to obtain $(x,y)$ is not needed --- the expected numbers 
of events $\langle N_i\rangle$ as functions of $(x,y)$ can be included 
in the likelihood. Thus the variables $(x,y)$ become free 
parameters of the combined likelihood fit, and the assumption that the 
number of signal events has a Gaussian distribution is not needed. 

Both approaches are tested with  
the control samples and MC simulation. We choose the combined fit 
approach as the baseline, but the procedure with separate fits in bins is 
also used: it allows a clear demonstration of the $CP$ asymmetry in each bin. 

\section{Event selection}

We use a data sample of $772\times 10^6$ $B\overline{B}$ pairs 
collected by the Belle detector. The decays \bdk\ and \bdpi\ 
are selected for the analysis. The neutral $D$ meson is reconstructed 
in the $K^0_S\pi^+\pi^-$ final state in all cases. We also select \dsdpi\ decays 
produced via the $e^+e^-\to c\bar{c}$ continuum process as 
a high-statistics sample to determine the $K_i$ parameters related to the 
flavor-tagged \dnkpp decay. 

The Belle detector is described in detail elsewhere \cite{belle,svd2}. 
It is a large-solid-angle magnetic spectrometer consisting of a
silicon vertex detector (SVD), a 50-layer central drift chamber (CDC) for
charged particle tracking and specific ionization measurement ($dE/dx$), 
an array of aerogel threshold Cherenkov counters (ACC), time-of-flight
scintillation counters (TOF), and an array of CsI(Tl) crystals for 
electromagnetic calorimetry (ECL) located inside a superconducting solenoid coil
that provides a 1.5 T magnetic field. An iron flux return located outside 
the coil is instrumented to detect $K_L$ mesons and identify muons (KLM).

Charged tracks are required to satisfy criteria based on the quality of the 
track fit and the distance from the interaction point of the beams (IP). 
We require each track to have a transverse momentum greater than 100 MeV/$c$, 
and the impact parameter relative to the IP to be less than 2 mm in 
the transverse and less than 10 mm in longitudinal projections. 
Separation of kaons and pions is accomplished by combining the responses of 
the ACC and the TOF with the $dE/dx$ measurement from the CDC. 
Neutral kaons are reconstructed from pairs of oppositely charged tracks
with an invariant mass $M_{\pi\pi}$ within $7$ MeV/$c^2$ of the nominal 
$K^0_S$ mass, flight distance from the IP in the plane transverse to 
the beam axis greater than 0.1 mm, 
and the cosine of the angle between the projections of $K^0_S$ 
flight direction and its momentum greater than 0.95. 

The flavor of the neutral $D$ mesons used for $K_i$ determination 
is tagged by the charge of the slow 
pion in the decay \dsdpi. The slow pion track is required to originate 
from the $D^0$ decay vertex to improve the momentum and angular resolution. 
The selection of signal candidates is based on two variables, 
the invariant mass of the neutral $D$ candidates $M_D=M_{K^0_S\pi^+\pi^-}$ 
and the difference of the invariant masses of the $D^{*\pm}$ and the 
neutral $D$ candidates 
$\Delta M=M_{(K^0_S\pi^+\pi^-)_D\pi}-M_{K^0_S\pi^+\pi^-}$. 
We retain the events satisfying the following criteria: 
$1800\mbox{ MeV}/c^2<M_D<1920$ MeV/$c^2$ and $\Delta M<150$ MeV/$c^2$.
We also require the momentum of the $D^0$ candidate in the CM frame 
$p_D$ to be in the range $1.8\mbox{ GeV}/c<p_D<2.8\mbox{ GeV}/c$
to reduce the effect 
of the efficiency profile on the $\phi_3$ measurement (see Sec.~\ref{sec:eff}). 
About 15\% of selected events contain more than one $D^{*\pm}$ candidate
that satisfies the requirements above; in this case we keep only one 
randomly selected candidate. 

Selection of \bdk and \bdpi samples is based on the CM-energy difference
$\de = \sum E_i - E_{\rm beam}$ and the beam-constrained $B$ meson mass
$\mbc = \sqrt{E_{\rm beam}^2 - (\sum \vec{p}_i)^2}$, where $E_{\rm beam}$ 
is the CM beam energy, and $E_i$ and $\vec{p}_i$ are the CM energies and 
momenta of the $B$ candidate decay products. We select events with 
$\mbc>5.2$ GeV/$c^2$ and $|\de|<0.18$~GeV for further analysis. 
We also impose a requirement on the invariant mass of the neutral $D$ 
candidate $|M_{\kspp}-M_{D^0}|<11$~MeV/$c^2$. 

Further separation of the background from $e^+e^-\to q\bar{q}$ ($q=u, d, s, c$) 
continuum events is done by calculating two variables that characterize the 
event shape. One is the cosine of the thrust angle \thr, 
where $\theta_{\rm thr}$ is the angle between the thrust axis of 
the $B$ candidate daughters and that of the rest of the event, 
calculated in the CM frame. 
The other is a Fisher discriminant \fish composed of 11 parameters \cite{fisher}: 
the production angle of the $B$ candidate, the angle of the $B$ thrust 
axis relative to the beam axis, and nine parameters representing 
the momentum flow in the event relative to the $B$ thrust axis in the CM frame.
We use the \de, \mbc, \thr, and \fish variables in the maximum likelihood fit. 

In both flavor $D^0$ and \bdk (\bdpi) samples, the momenta of the tracks 
forming a $D^0$ candidate are constrained to give the nominal $D^0$ mass in 
the calculation of the Dalitz plot variables. 

\section{Flavor-tagged sample \dsdpi, \dkpp}

\label{sec:flavor}

The number of events $K_i$ in bin $i$ of the flavor-tagged \dkpp decay is  
obtained from a two-dimensional unbinned fit to the distribution 
of $M_{D}$ and $\Delta M$ variables. The fits in each Dalitz plot bin 
are performed independently. 
The fit uses a signal PDF and two background components: purely 
random combinatorial background
and background with a real $D^0$ and random slow pion track. 
The signal distribution is a product of the PDFs for $M_D$ (triple Gaussian) and 
$\Delta M$ (sum of bifurcated Student's $t$-distribution and bifurcated 
Gaussian distribution). The combinatorial background is parameterized 
by a linear function in $M_D$ and by a function 
with a kinematic threshold at the $\pi^+$ mass in $\Delta M$: 
\begin{equation}
\begin{split}
  p_{\rm comb}(\Delta M)&=\sqrt{y}(1+Ay[1+B(M_D-m_{D^0})])e^{-yC}\\
\end{split}
\label{eq:bck_comb}
\end{equation}
where $y=\Delta M-m_{\pi^+}$, $m_{\pi^+}$ and $m_{D^0}$ are the nominal 
masses of $\pi^+$ and $D^0$, respectively, and $A$, $B$, and $C$ are free 
parameters. A small correlation between the $M_D$
and $\Delta M$ distributions is introduced that is controlled by the 
parameter $B$. The random slow pion background is parameterized as a 
product of the signal $M_D$ distribution and combinatorial $\Delta M$ 
background shape. 

The parameters of the signal and background distributions are obtained 
from the fit to data. The parameters of the signal PDF are constrained 
to be the same in all bins. The free parameters in each bin are 
the number of signal events $K_i$, the parameters of the 
background distribution, and fractions of the background components. 

\begin{figure}
  \includegraphics[width=0.5\textwidth]{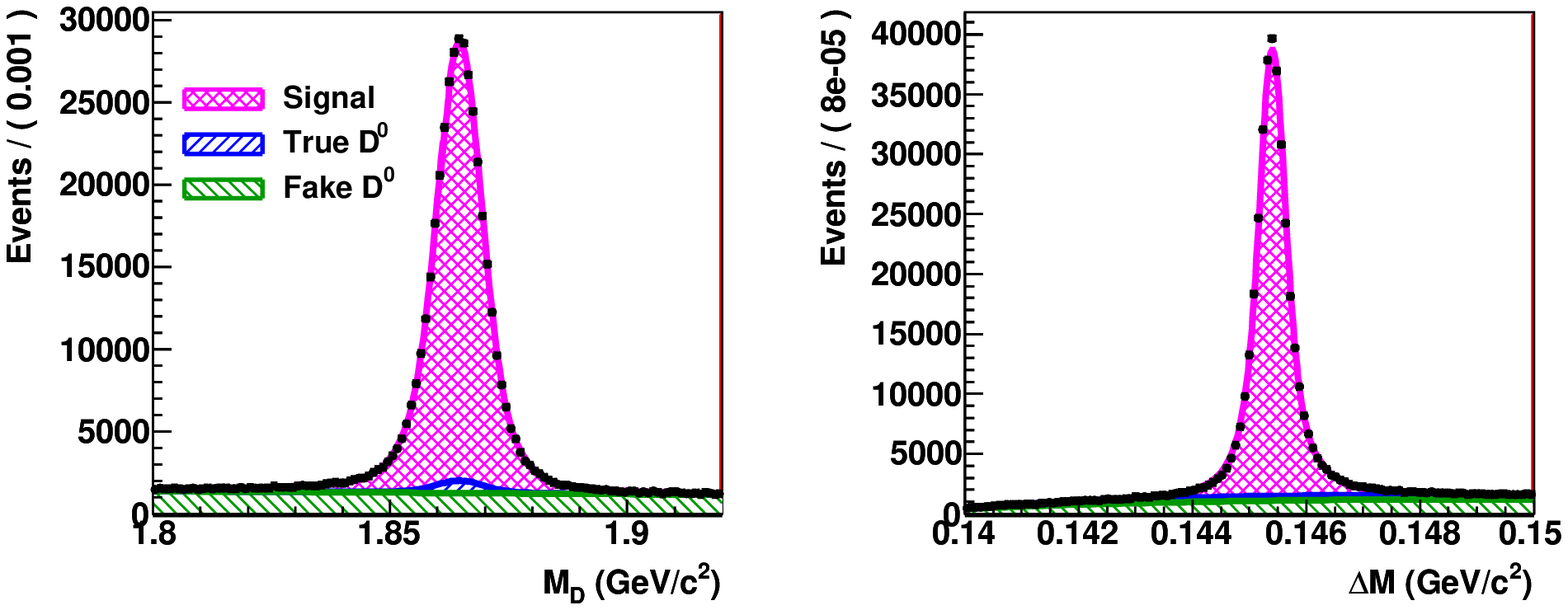}
  \put(-153,85){(a)}
  \put(-25,85){(b)}
  \caption{Projections of the flavor-tagged \dsdpi, \dkpp data with 
           $1.8\mbox{ GeV}/c<p_D<2.8\mbox{ GeV}/c$. 
           (a) $M_D$ distribution for $144.5$ MeV/$c^2<\Delta M<146.5$ MeV/$c^2$. 
           (b) $\Delta M$ distribution for $1854$ MeV/$c^2<M_D<1876$ MeV/$c^2$.
           Histograms show the fitted signal and background 
           contributions, points with the error bars are the data.
           The full \dkpp Dalitz plot is used. }
  \label{fig:dsdpi_dmmd}
\end{figure}

The fit results from the flavor-tagged $D$ sample integrated over the
whole Dalitz plot are shown in Fig.~\ref{fig:dsdpi_dmmd}. The number of signal events 
calculated from the integral of the signal distribution is 
$426938\pm 825$, the background fraction in the signal region 
$|M_D-m_{D^0}|<11$ MeV/$c^2$, $144.5\mbox{ MeV}/c^2<\Delta M<146.5$ MeV/$c^2$
is $10.1\pm 0.1$\%. The signal yield in bins is shown in 
Table~\ref{tab:flavor_bins}. 

\begin{table}
  \caption{Signal yields in Dalitz plot bins for the flavor-tagged 
  \dsdpi, \dkpp sample with $1.8\mbox{ GeV}/c<p_D<2.8\mbox{ GeV}/c$. }
  \label{tab:flavor_bins}
  \begin{tabular}{|r|c|c|}
   \hline
    Bin $i$  & $K_i$  & $K_{-i}$ \\
   \hline
    1 & $43261\pm 255$ & \phantom{$0$}$8770\pm 124$  \\
    2 & $58005\pm 268$ & $1827\pm 63$                \\
    3 & $62808\pm 274$ & $1601\pm 58$                \\
    4 & $44513\pm 253$ & $26482\pm 202$              \\
    5 & $21886\pm 177$ & $13146\pm 143$              \\
    6 & $28876\pm 197$ & $1765\pm 68$                \\
    7 & $48001\pm 265$ & $22476\pm 196$              \\
    8 & \phantom{$0$}$9279\pm 125$ & $26450\pm 181$  \\
   \hline
    Total & \multicolumn{2}{c|}{$426938\pm 825$\phantom{$0$}} \\
   \hline
  \end{tabular}
\end{table}

\section{Selection of \bdpi and \bdk samples}

The decays \bdk and \bdpi have similar topology and background 
sources and their selection is performed in a similar way. 
The mode \bdpi has an order of magnitude larger branching 
ratio and a smaller amplitude ratio $r_B\sim 0.01$ due to the ratio 
of weak coefficients 
$|V_{ub}^{\vphantom{*}} V^*_{cd}|/
 |V_{cb}^{\vphantom{*}} V^*_{ud}|\sim 0.02$ and the color 
suppression factor. This results in the small $CP$ violation in this mode, 
therefore it is used as a control sample to test the procedures of the background extraction
and Dalitz plot fit. In addition, signal resolutions in \de and \mbc and the 
Dalitz plot structure of some background components are constrained 
from the control sample and used in the signal fit. 

The number of signal events is obtained by fitting 
the 4D distribution of variables \mbc, \de, \thr and \fish. 
The fits to the \bdpi and \bdk samples use the following three background 
components in addition to the signal PDF:
\begin{itemize}
  \item Combinatorial background from the process $e^+e^-\to q\overline{q}$, 
        where $q=(u,d,s,c)$. 
  \item Random $B\overline{B}$ background, in which the tracks forming the 
        \bdpi candidate come from decays of both $B$ mesons in the event. 
        The number of possible $B$ decay combinations that contribute 
        to this background is large, therefore both the Dalitz 
        distribution and $(\mbc, \de)$ distribution are quite smooth. 
  \item Peaking $B\overline{B}$ background, in which all tracks forming 
        the \bdpi candidate come from the same $B$ meson.  This kind of 
        background is dominated by $B\to D^*\pi$ decays 
        reconstructed without the $\pi$ or $\gamma$ from the 
        $D^{*}$ decay. 
\end{itemize}
In addition, the \bdk fit includes 
a fourth component that models \bdpi decays in which
the pion is misidentified as a kaon. 

The PDF for the signal parameterization (as well as for each of the 
background components) is a product of the $(\mbc,\de)$ and $(\thr,\fish)$
PDFs. The $(\mbc,\de)$ PDF is a 
2D double-Gaussian function, which has a correlation between \mbc and \de. The 
double-Gaussian function models both the core and tails of the distribution.
The $(\thr,\fish)$ distribution is 
parameterized by the sum of two functions (with different coefficients) 
of the form
\begin{equation}
  \begin{split}
  p(x,\fish) = & \exp(C_1x+C_2x^2+C_3x^3)\times \\
               & G(\fish, F_0(x), \sigma_{FL}(x), \sigma_{FR}(x)), 
  \end{split}
  \label{thrfish_param}
\end{equation}
where $x=\thr$, $G(\fish,F,\sigma_L,\sigma_R)$ is the bifurcated 
Gaussian distribution with the mean $F$ and the widths $\sigma_L$
and $\sigma_R$, and functions $F_0$, $\sigma_{FL}$ and $\sigma_{FR}$
are polynomials that contain only even powers of $x$. 
The parameters of the signal PDF are obtained from the signal MC 
simulation. However, to account for the possible imperfection of 
the simulation, we allow all the width parameters to scale by a 
common factor, which is obtained from the \bdpi sample.

The combinatorial background from continuum $e^+e^-\to q\overline{q}$
production is obtained from the experimental sample collected at a 
CM energy below the $\Upsilon(4S)$ resonance (off-resonance data). 
The parameterization in variables $(\thr,\fish)$ follows 
Eq.~(\ref{thrfish_param}). 
The parameterization in $(\mbc,\de)$ is the product of an exponential 
distribution in $\Delta E$ and the empirical shape proposed by the 
ARGUS collaboration~\cite{argus} in $\mbc$: 
\begin{equation}
  p_{\rm comb}(\mbc, \de)=\exp(-\alpha\de)\mbc\sqrt{y}\exp(-cy), 
\end{equation}
where $y=1-\mbc/E_{\rm beam}$, $E_{\rm beam}$ is the CM beam energy, 
and $\alpha$ and $c$ are empirical parameters. 

The parameters for random and peaking $B\overline{B}$ backgrounds are obtained from 
a generic MC sample. Generator information is used to distinguish between 
the two: the latter contains only the events in which the candidate 
is formed of tracks coming from both $B$ mesons. The $(\mbc,\de)$
distributions for each of these backgrounds are parameterized by 
the sum of three components:
\begin{itemize}
  \item the product of an exponential (in $\de$) and Argus (in $\mbc$) 
  functions, as for continuum background (as expected, 
  this component dominates the random $B\overline{B}$ background); 
  \item the product of an exponential in the $\de$ and bifurcated Gaussian 
  distribution in $\mbc$, where the mean of the Gaussian distribution 
  is linear as a function of $\de$; and 
  \item a two-dimensional Gaussian distribution in $\de$ and $\mbc$, 
  which includes a correlation and is asymmetric in $\mbc$. 
  This component is small compared to the 
  random $B\overline{B}$ contribution, but dominates 
  the peaking $B\overline{B}$ background, which mostly consists of 
  partially reconstructed $B$ decays. 
\end{itemize}

The peaking background coming from $B^+B^-$ and $B^0\overline{B}{}^0$ decays is treated 
separately in $(\mbc,\de)$ variables, while a common 
$(\thr,\fish)$ distribution is used. In the case of the \bdk fit, 
\bdpi events with the pion misidentified as a kaon are treated as a
separate background category. The distributions of $\mbc, \de$
and $\thr, \fish$ variables are parameterized in the same way as for
the signal events and are obtained from MC simulation. 

The Dalitz plot distributions of the background components are 
discussed in the next section. Note that the Dalitz distribution is 
described by the relative number of events in each bin. The numbers of 
events in bins can be free parameters in the fit, thus 
there will be no uncertainty due to 
the modeling of the background distribution over the Dalitz plot
in such an approach. This procedure is justified for 
background that is either well separated from the signal (such as peaking 
$B\overline{B}$ background in the case of \bdpi), or 
is constrained by a much larger number of events than the signal (such 
as the continuum background). 

\begin{figure}
  \includegraphics[width=0.5\textwidth]{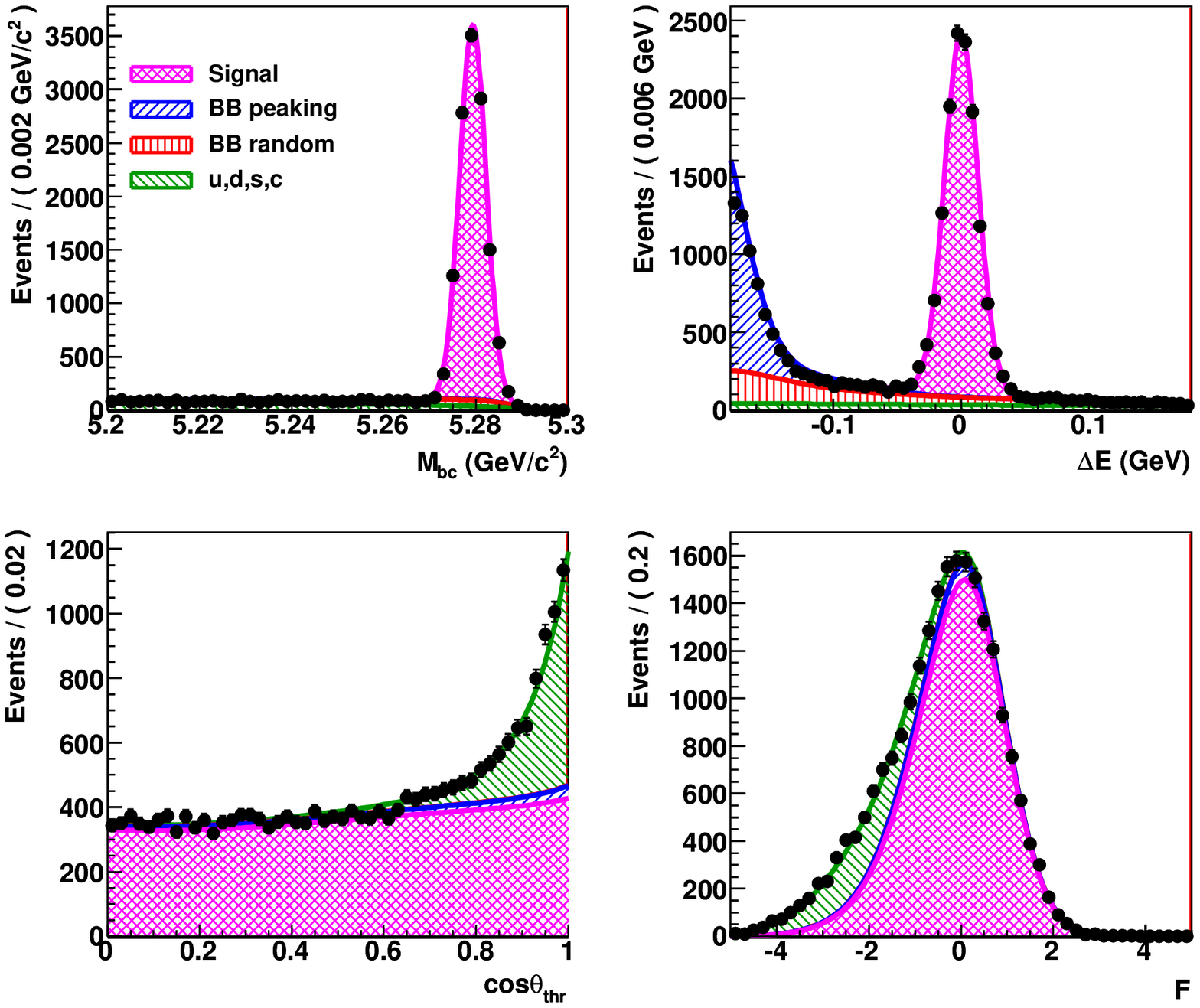}
  \put(-152,197){(a)}
  \put(-25,197){(b)}
  \put(-152,90){(c)}
  \put(-25,90){(d)}
  \caption{Projections of the \bdpi data. 
           (a) \mbc distribution with $|\de|<30$ MeV and $\thr<0.8$ requirements. 
           (b) \de distribution with $\mbc>5270$ MeV/$c^2$ and $\thr<0.8$ requirements. 
           (c) \thr and (d) \fish distributions with $|\de|<30$ MeV and $\mbc>5270$ MeV/$c^2$
           requirements.
           Histograms show the fitted signal and background 
           contributions, points with error bars are the data. 
           The entire \dkpp Dalitz plot is used. }
  \label{fig:bdpi_exp_all}
\end{figure}

\begin{figure}
  \includegraphics[width=0.5\textwidth]{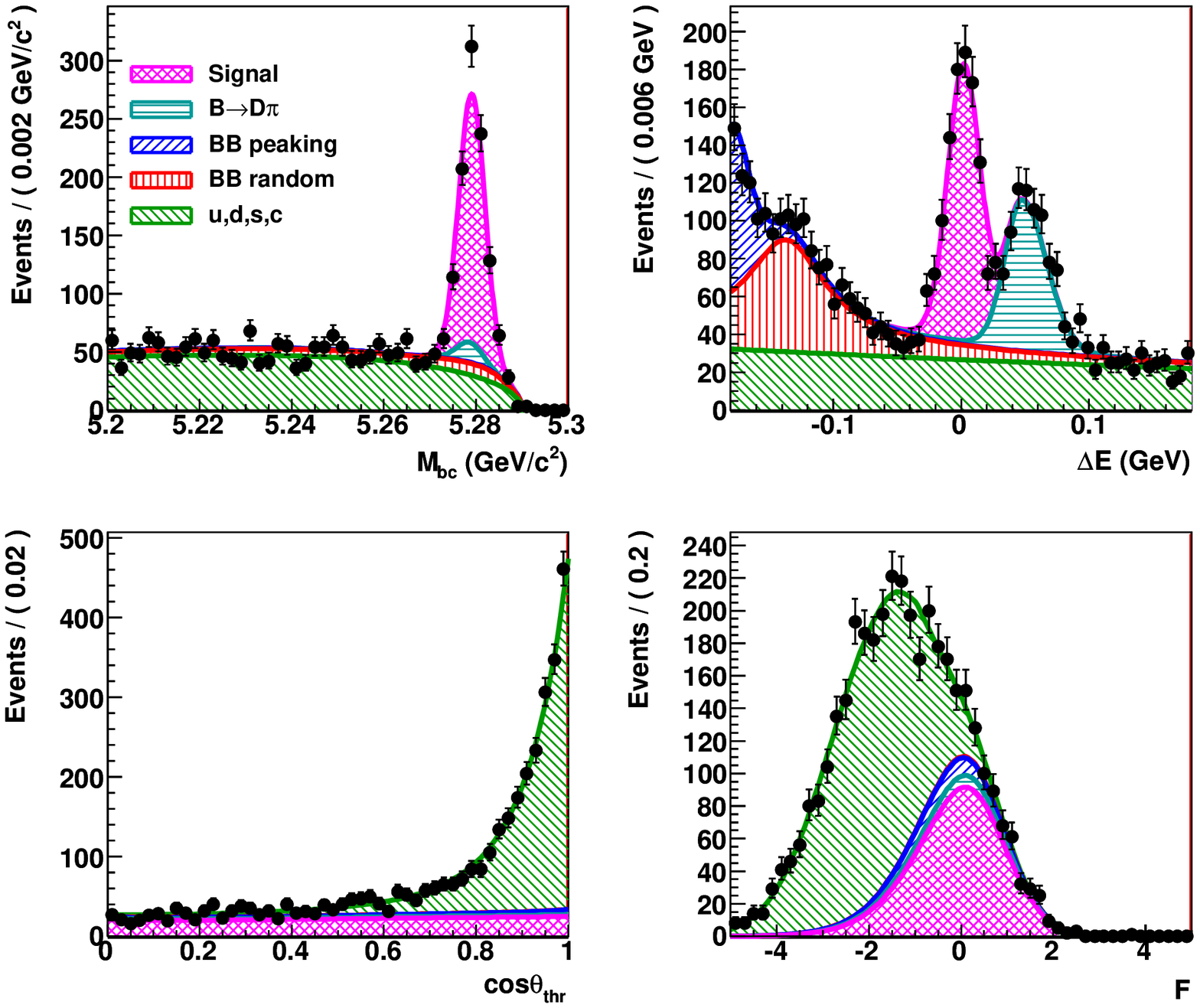}
  \put(-152,197){(a)}
  \put(-25,197){(b)}
  \put(-152,90){(c)}
  \put(-25,90){(d)}
  \caption{Projections of the \bdk data. 
           (a) \mbc distribution with $|\de|<30$ MeV and $\thr<0.8$ requirements. 
           (b) \de distribution with $\mbc>5270$ MeV/$c^2$ and $\thr<0.8$ requirements. 
           (c) \thr and (d) \fish distributions with $|\de|<30$ MeV and $\mbc>5270$ MeV/$c^2$
           requirements. 
           Histograms show the fitted signal and background 
           contributions, points with error bars are the data. 
           The entire \dkpp Dalitz plot is used. }
  \label{fig:bdk_exp_all}
\end{figure}

The results of the fit to \bdpi and \bdk data with the full Dalitz plot 
taken are shown in Figs.~\ref{fig:bdpi_exp_all} and \ref{fig:bdk_exp_all}, 
respectively. We obtain a total of $19106\pm 147$ signal \bdpi events and 
$1176\pm 43$ signal \bdk events --- 55\% more than in the 
605 fb$^{-1}$ model-dependent analysis~\cite{belle_phi3_3}. 
The improvement partially comes from the larger integrated luminosity of the sample, 
and partially from the larger selection efficiency due to improved track 
reconstruction. 

\section{Data fits in bins}

The data fits in bins for both \bdpi and \bdk samples are performed with two 
different procedures: separate fits for the number of events in bins
and the combined fit with the free parameters $(x,y)$ as discussed in 
Sec.~\ref{sec:fit_procedure}. The combined fit is used to obtain
the final values for $(x,y)$, while the separate fits provide a 
crosscheck of the fit procedure and a way to visualize the 
extent of $CP$ violation within the sample.
A study with MC pseudo-experiments is performed 
to check that the observed difference in the fit results 
between the two approaches agrees with expectation. 

In the case of separate fits in bins, we first perform the fit to all 
events in the Dalitz plot. The fit uses background shapes fixed 
to those obtained from fits to the
generic MC samples of continuum and $B\overline{B}$ decays. 
The signal shape 
parameters are fixed to those obtained from a fit to the signal 
MC sample except for the mean value and width scale factors of \de and \mbc 
PDFs. 
As a next step, we fit the 4D $(\mbc, \de, \thr, \fish)$ distributions
in each bin separately, with the signal peak positions and 
width scale factors fixed to the values obtained from the fit to all events. 
The free parameters of each fit are the number of signal events, 
and the number of events in each background category. 

\begin{figure}
  \includegraphics[width=0.5\textwidth]{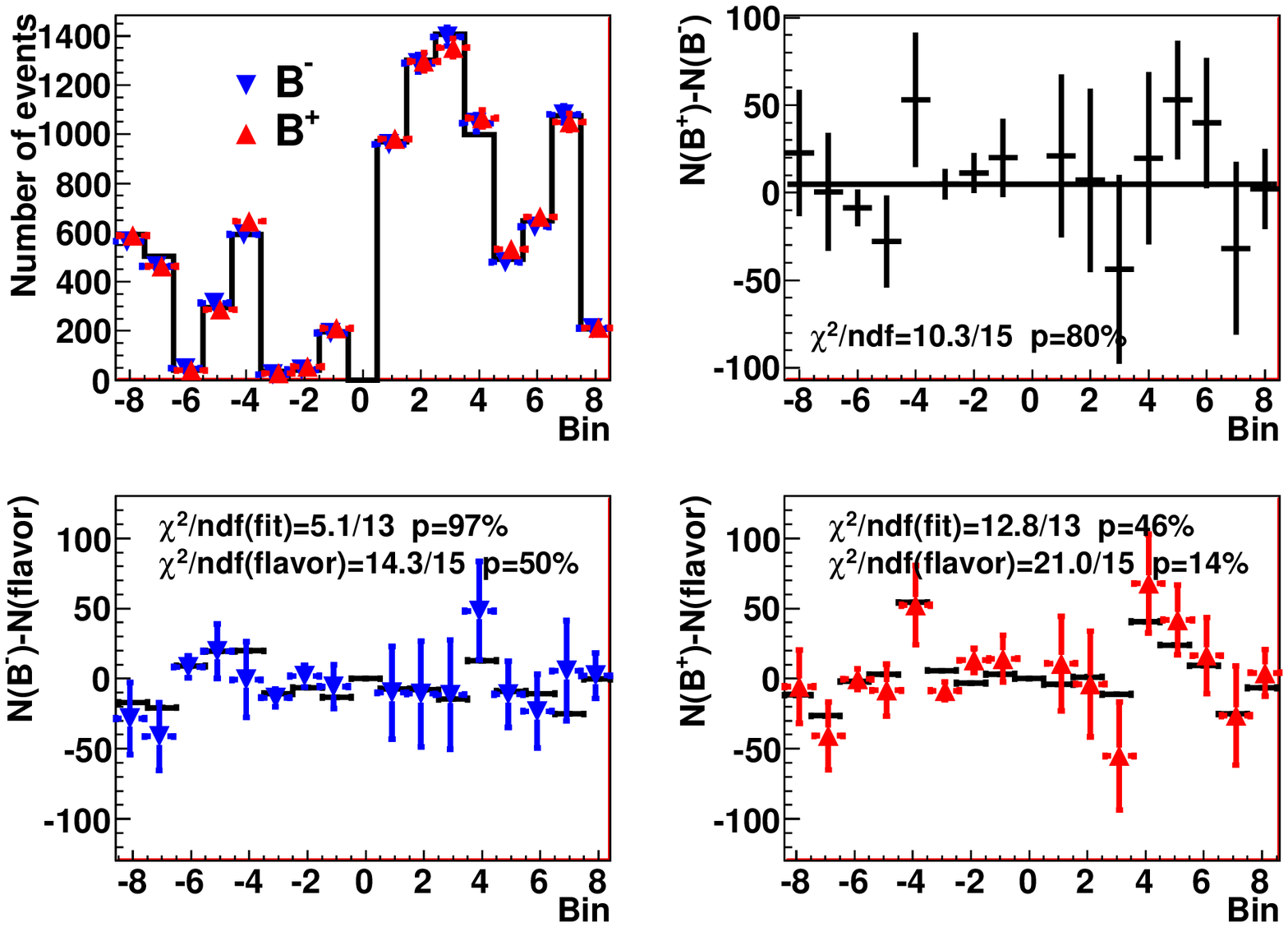}
  \put(-227,165){(a)}
  \put(-99,165){(b)}
  \put(-227,22){(c)}
  \put(-99,22){(d)}
  \caption{Results of the fit to the \bdpi sample. 
  (a)~Signal yield in bins of the \dkpp Dalitz plot: from \bdpim (red), 
  \bdpip (blue) and flavor sample (histogram). 
  (b)~Difference of signal yields between the \bdpip and \bdpim decays. 
  (c)~Difference of signal yields between the \bdpim and flavor samples
  (normalized to the total \bdpim yield): yield from the separate fits (points with 
  error bars), and as a result of the combined $(x,y)$ fit (horizontal bars).
  (d)~Same as (c) for \bdpip data. }
  \label{fig:bdpi_bin_n}
\end{figure}

The numbers of signal events in bins for the \bdpi sample extracted from the fits are given in 
Table~\ref{tab:bdpi_signal_num}. These numbers are used in the fit to extract 
$(x,y)$ using Eq.~(\ref{n_b}) after the cross-feed and efficiency correction for 
both $N_i$ and $K_i$. Figure~\ref{fig:bdpi_bin_n} illustrates the results of
this fit. The numbers of signal events in each bin for $B^+$ and $B^-$
are shown in Fig.~\ref{fig:bdpi_bin_n}(a) together with the numbers of events in the
flavor-tagged $D^0$ sample (appropriately scaled). The difference in the number of signal events 
shown in Fig.~\ref{fig:bdpi_bin_n}(b) does not reveal $CP$ violation. 
Figures~\ref{fig:bdpi_bin_n}(c) [(d)] show the difference between the numbers of 
signal events for $B^+$ [$B^-$] data and scaled flavor-tagged $D^0$ sample, both for the 
data and after the $(x,y)$ fit. The $\chi^2/\mbox{ndf}$ is reasonable for both the 
$(x,y)$ fit and the comparison with the flavor-specific $CP$ conserving amplitude. 

\begin{figure}
  \includegraphics[width=0.5\textwidth]{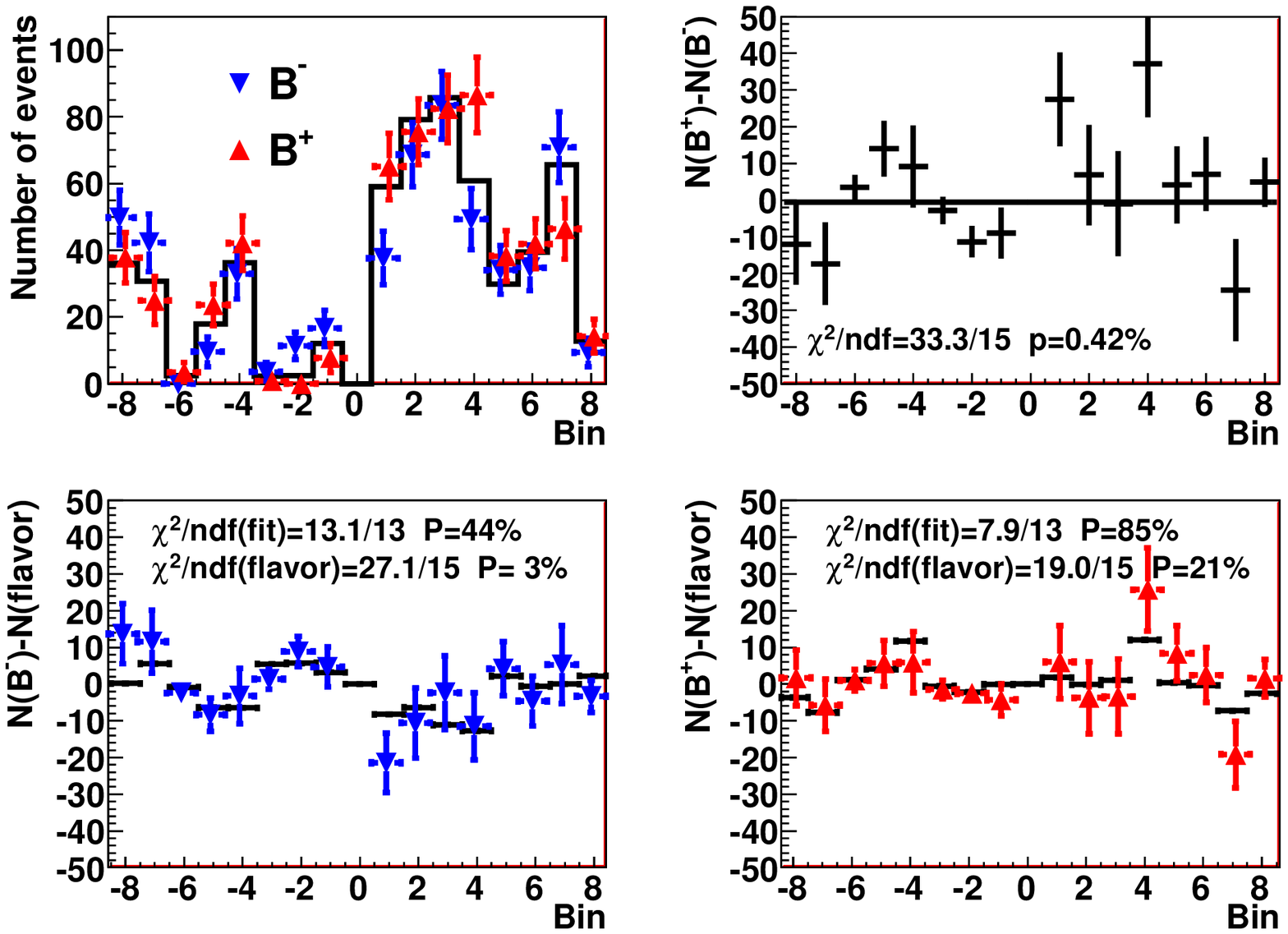}
  \put(-227,165){(a)}
  \put(-99,165){(b)}
  \put(-227,22){(c)}
  \put(-99,22){(d)}
  \caption{Results of the fit of the \bdk sample. 
  (a)~Signal yield in bins of the \dkpp Dalitz plot: from \bdkm (red), 
  \bdkp (blue) and flavor sample (histogram). 
  (b)~Difference of signal yields between the \bdkp and \bdkm decays. 
  (c)~Difference of signal yields between the \bdkm and flavor samples
  (normalized to the total \bdkm yield): yield from the separate fits (points with 
  error bars), and as a result of the combined $(x,y)$ fit (horizontal bars).
  (d)~Same as (c) for \bdkp data. }
  \label{fig:bdk_bin_n}
\end{figure}

Unlike \bdpi, the \bdk sample has significantly different 
signal yields in bins of $B^+$ and $B^-$ data (see 
Fig.~\ref{fig:bdk_bin_n}(b) and Table~\ref{tab:bdk_signal_num}). 
The probability to obtain this difference 
as a result of a statistical fluctuation is 0.42\%. This value can be 
taken as the model-independent measure of the $CP$ violation significance.
The significance of $\phi_3$ being nonzero is in general 
smaller since $\phi_3\neq 0$ results in a specific pattern of charge 
asymmetry. The fit of the signal yields to the expected 
pattern described by the parameters $(x,y)$ is of good 
quality~[Figs. \ref{fig:bdk_bin_n}(c,d)], which is consistent 
with the hypothesis that the observed $CP$ violation is solely 
explained by the mechanism involving nonzero $\phi_3$. 

\begin{table}
  \caption{Signal yields in Dalitz plot bins for the \bdpi, 
  \dkpp sample with the optimal binning. }
  \label{tab:bdpi_signal_num}
  \begin{tabular}{|r|c|c|}
   \hline
    Bin $i$  & $N^{-}_i$ & $N^{+}_i$ \\
   \hline
   -8     &  $564.2\pm 25.3$             &  $587.0\pm 25.7$             \\
   -7     &  $462.3\pm 23.8$             &  $462.8\pm 23.9$             \\
   -6     &   $47.9\pm 7.7$              &   $39.2\pm 7.2$              \\
   -5     &  $314.1\pm 19.0$             &  $286.2\pm 18.2$             \\
   -4     &  $592.6\pm 26.5$             &  $645.7\pm 27.8$             \\
   -3     &   $22.2\pm 6.2$              &   $27.2\pm 6.3$              \\
   -2     &   $42.7\pm 7.6$              &   $54.0\pm 8.7$              \\
   -1     &  $190.8\pm 15.4$             &  $210.8\pm 16.3$             \\
    1     &  $959.2\pm 32.6$             &  $980.2\pm 33.1$             \\
    2     & $1288.7\pm 37.0$\phantom{$0$}& $1295.9\pm 37.1$\phantom{$0$}\\
    3     & $1395.8\pm 38.4$\phantom{$0$}& $1352.2\pm 37.9$\phantom{$0$}\\
    4     & $1045.5\pm 34.7$\phantom{$0$}& $1065.1\pm 34.9$\phantom{$0$}\\
    5     &  $479.3\pm 23.3$             &  $532.2\pm 24.5$             \\
    6     &  $623.7\pm 26.0$             &  $663.5\pm 26.7$             \\
    7     & $1081.0\pm 35.3$\phantom{$0$}& $1049.2\pm 34.8$\phantom{$0$}\\
    8     &  $210.0\pm 16.1$             &  $212.1\pm 16.3$             \\
   \hline
    Total & $9467.1\pm 103.6$            & $9639.1\pm 104.7$            \\
   \hline
  \end{tabular}
\end{table}

\begin{table}
  \caption{Signal yields in Dalitz plot bins for the \bdk, 
  \dkpp sample with the optimal binning. }
  \label{tab:bdk_signal_num}
  \begin{tabular}{|r|c|c|}
   \hline
    Bin $i$  & $N^{-}_i$ & $N^{+}_i$ \\
   \hline
   -8     &  $49.8\pm 8.2$              &  $37.8\pm 7.5$             \\
   -7     &  $42.2\pm 8.6$              &  $24.9\pm 7.2$             \\
   -6     &\phantom{$0$}$0.0\pm 1.9$    & \phantom{$0$}$3.4\pm 2.9$  \\
   -5     &\phantom{$0$}$9.6\pm 4.5$    &  $23.6\pm 6.2$             \\
   -4     &  $32.9\pm 7.5$              &  $42.1\pm 8.3$             \\
   -3     &\phantom{$0$}$3.5\pm 2.8$    & \phantom{$0$}$0.7\pm 2.5$  \\
   -2     &   $11.3\pm 4.1$             & \phantom{$0$}$0.0\pm 1.3$  \\
   -1     &  $16.6\pm 5.4$              & \phantom{$0$}$7.7\pm 4.4$  \\
    1     &  $37.6\pm 8.0$              &  $65.1\pm 9.9$             \\
    2     & $68.6\pm 9.6$               & $75.5\pm 9.8$              \\
    3     &\phantom{$0$}$83.4\pm 10.1$  & \phantom{$0$}$82.4\pm 10.2$ \\
    4     & $49.3\pm 9.1$               & \phantom{$0$}$86.5\pm 11.4$ \\
    5     &  $34.0\pm 7.3$              &  $38.3\pm 7.6$             \\
    6     &  $34.8\pm 6.8$              &  $41.9\pm 7.5$             \\
    7     &\phantom{$0$}$70.8\pm 10.6$  & $46.4\pm 9.0$              \\
    8     &\phantom{$0$}$9.4\pm 4.3$    &  $14.2\pm 5.1$             \\
   \hline
    Total & $574.9\pm 29.9$             & $601.6\pm 30.8$            \\
   \hline
  \end{tabular}
\end{table}

\begin{figure}
  \begin{center}
  \includegraphics[width=0.37\textwidth]{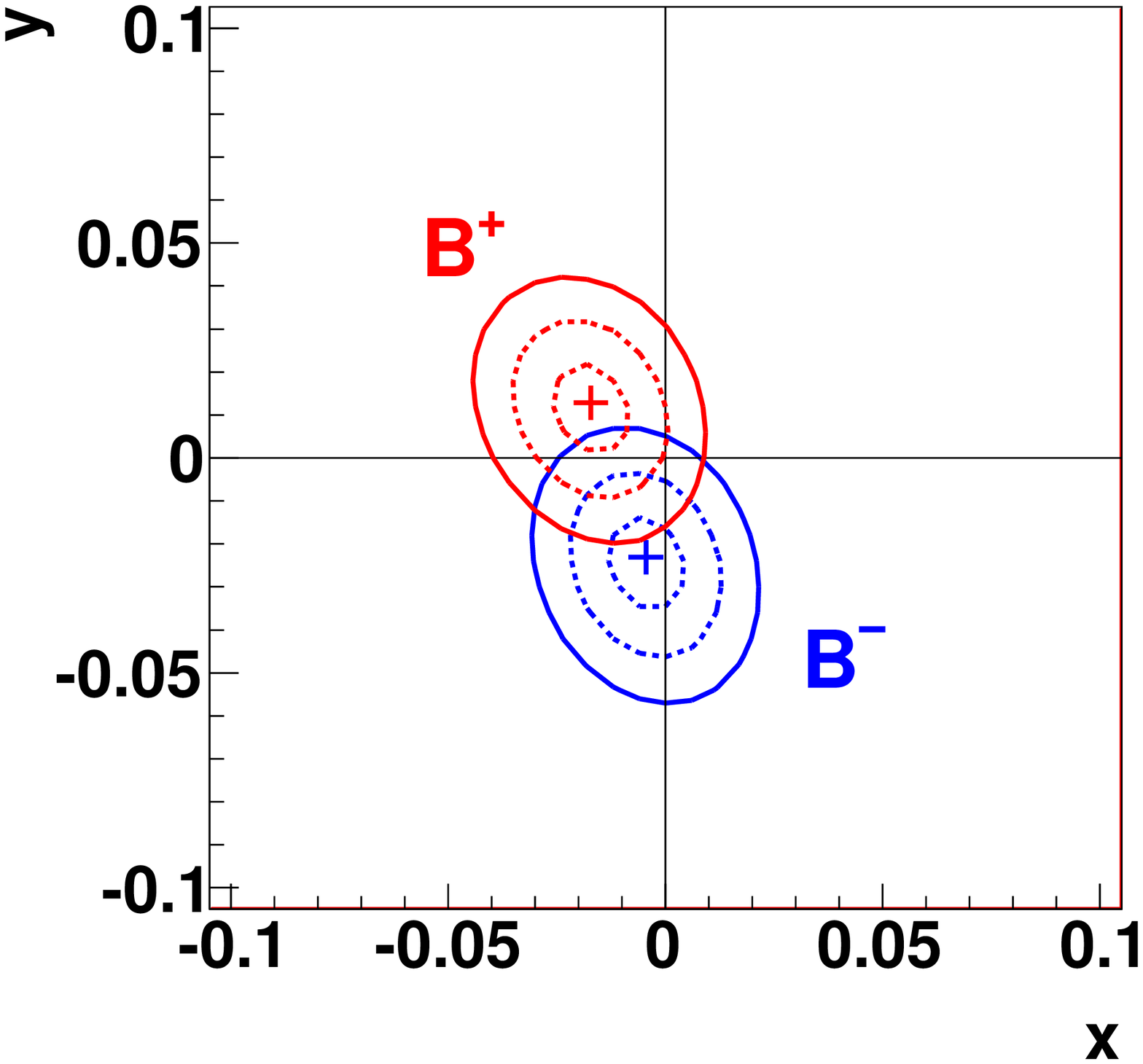}
  \put(-30,150){(a)}

  \includegraphics[width=0.37\textwidth]{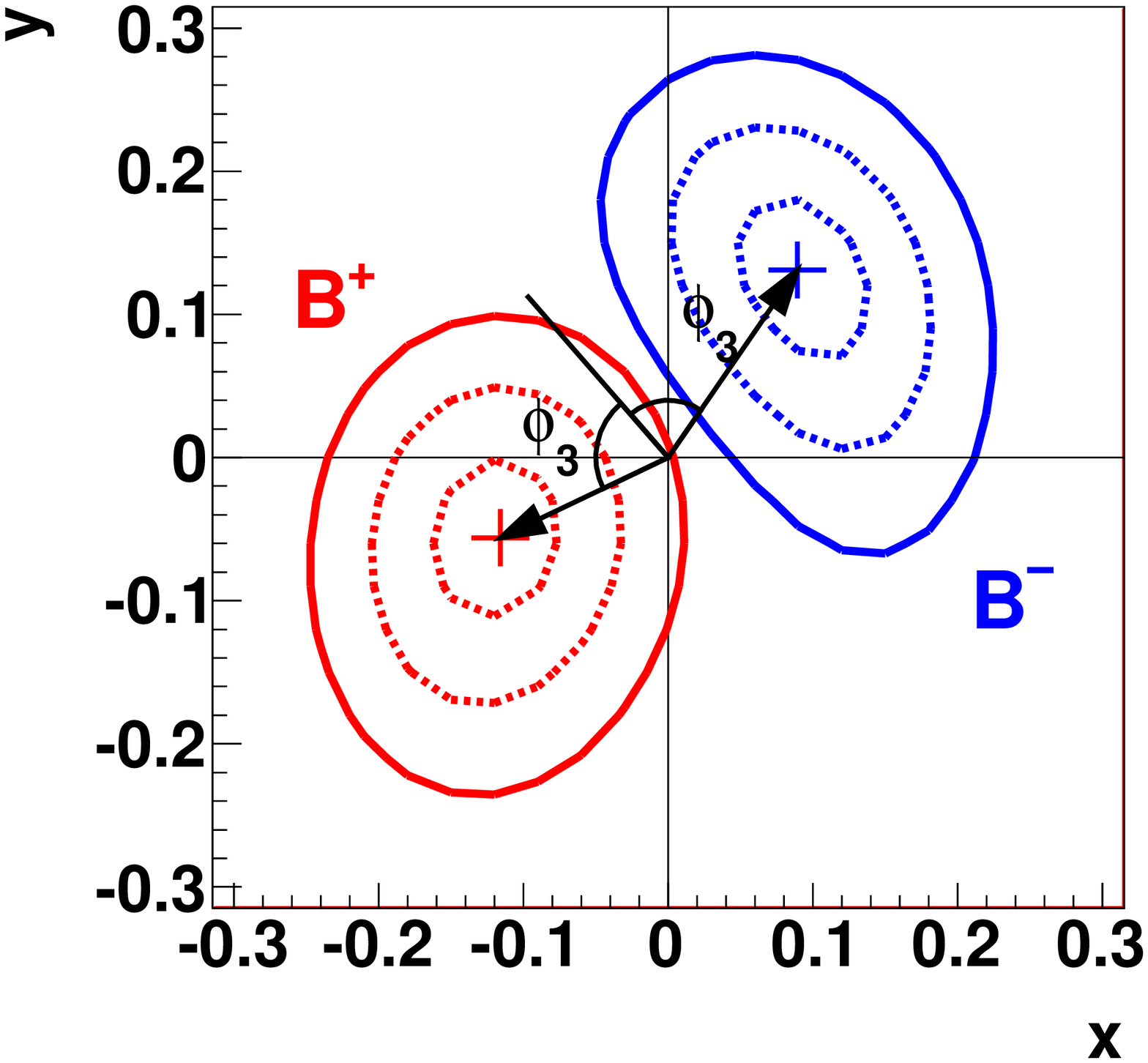}
  \put(-30,150){(b)}
  \end{center}
  \caption{One, two, and three standard deviation $(x,y)$ 
           confidence levels for (a) \bdpi and (b) \bdk decays (statistical only). 
           Note different scale for \bdpi and \bdk modes. 
           The weak phase $\phi_3$ appears as half
           the opening angle between ($x_+, y_+$) and ($x_-, y_-$)
           vectors. }
  \label{fig:bdk_xy}
\end{figure}

The default combined fit constrains the random $B\overline{B}$
background in bins from the generic MC, and takes the $(x_{\pm},y_{\pm})$ variables as free 
parameters. Fits to $B^+$ and $B^-$ data are performed separately. 
The plots illustrating the combined fit results are given in the Appendix.
Additional free parameters are the yields of continuum and peaking 
$B\overline{B}$ backgrounds in each bin, the fraction of the random $B\overline{B}$
background, and the means and scale factors of the signal \mbc and \de distributions. 
The values of $(x,y)$ are then corrected for the fit bias obtained from 
MC pseudo-experiments. The value of the bias depends on the initial 
$x$ and $y$ values and is of the order $5\times 10^{-3}$ for the \bdk sample 
and less than $10^{-3}$ for \bdpi sample. 

\begin{table*}
  \caption{$(x,y)$ parameters and their statistical correlations from the combined 
  fit of the \bdpi and \bdk samples. The first error is statistical, 
the second error is systematic, and the 
third error is due to the uncertainty on the 
$c_i$ and $s_i$ parameters. }
  \label{xy_table}
  \begin{tabular}{|l|c|c|}
    \hline
      Parameter & \bdpi & \bdk \\
    \hline
      $x_-$ & $-0.0045\pm 0.0087\pm 0.0049\pm 0.0026$ & $+0.095\pm 0.045\pm 0.014\pm 0.010$ \\
      $y_-$ & $-0.0231\pm 0.0107\pm 0.0041\pm 0.0065$ & $+0.137^{+0.053}_{-0.057}\pm 0.015\pm 0.023$ \\
      $\rm{corr}(x_-, y_-)$ & $-0.189$ & $-0.315$ \\
    \hline
      $x_+$ & $-0.0172\pm 0.0089\pm 0.0060\pm 0.0026$ & $-0.110\pm 0.043\pm 0.014\pm 0.007$ \\
      $y_+$ & $+0.0129\pm 0.0103\pm 0.0059\pm 0.0065$ & $-0.050^{+0.052}_{-0.055}\pm 0.011\pm 0.017$ \\
      $\rm{corr}(x_+, y_+)$ & $-0.205$ & $+0.059$ \\
    \hline
  \end{tabular}
\end{table*}

The values of $(x,y)$ parameters and their statistical correlations 
obtained from the combined fit for \bdpi control sample and 
\bdk sample are given in Table~\ref{xy_table}. The measured values of 
$(x_{\pm}, y_{\pm})$ for both samples with their statistical 
likelihood contours are shown in Fig.~\ref{fig:bdk_xy}. 

\section{Systematic errors}

Systematic errors in the $(x,y)$ fit are obtained for the default 
procedure of the combined fit with the optimal binning. 
The systematic errors are summarized in Table~\ref{tab:syst}. 

The uncertainty due to the signal shape used in the fit includes the 
sources listed below: 
\begin{enumerate}
  \item The choice of parameterization used to model the shape. 
        The corresponding uncertainty is estimated by using 
        the non-parametric Keys PDF function~\cite{keys} instead 
        of the parameterized distribution. 
  \item Correlation between the $(\mbc, \de)$ 
        and $(\thr, \fish)$ distributions. To estimate its effect, 
        we use a 4D binned histogram to describe the distribution. 
  \item The MC description of the (\thr, \fish) distribution. Its 
        effect is estimated by floating the parameters 
        of the distribution in the fit to the \bdpi control sample. 
  \item The dependence of the signal width on the Dalitz plot bin. 
        The uncertainty due to this effect is estimated 
        by performing the \bdpi fit 
        with the shape parameters floated separately for each bin, and then 
        using the results in the fit to \bdk data. 
\end{enumerate}
We do not assign an uncertainty due to the difference in $(\mbc, \de)$ 
shape between the MC and data since the width of the signal distribution 
is calibrated using \bdpi data. 

For the uncertainty due to the continuum background shape, we use the same four 
sources as considered for the signal distribution. The uncertainty 
due to the choice of parameterization is estimated by using 
the Keys PDF as an alternative. The effect of correlation between the 
$(\mbc, \de)$ and $(\thr, \fish)$ distributions is estimated
by using a distribution split into the sum of two components 
($u,d,s$ and charm contributions) with independent $(\mbc, \de)$ and 
$(\thr, \fish)$ shapes. The uncertainty due to the MC description of 
the $(\mbc, \de)$ and $(\thr, \fish)$ distributions is estimated by 
floating their parameters in the \bdpi fit. To estimate the effect 
of correlation of the shape with the Dalitz plot variables
we fit the shapes separately in each Dalitz plot bin. 

The uncertainties due to the shapes of random and peaking $B\overline{B}$
backgrounds are estimated differently for the \bdpi\ and \bdk\ samples. 
In the \bdpi case, the effect of the background shape uncertainty is estimated by performing a fit with 
$\de>-0.1$ GeV: this requirement rejects the peaking $B\overline{B}$ 
background and a large part of the random $B\overline{B}$ background. 
In the case of the \bdk\ sample, the uncertainty is estimated by performing 
an alternative fit with the $(\mbc, \de)$ and $(\thr, \fish)$ shapes 
taken from the \bdpi\ sample (shifted by 50 MeV in \de to account for 
difference in pion and kaon masses in the calculation of \de). 

In the case of the fit to the \bdk sample, the uncertainty due to the 
\bdpi background shape in the $(\thr, \fish)$ variables is estimated 
by taking the $(\thr, \fish)$ shape for signal events. 
The Dalitz plot distribution uncertainty is estimated by using the 
number of flavor-tagged events in bins, rather than the number of \bdpi
events used in the default fit. Uncertainties due to 
correlations are treated in the same way as in the case 
of the signal distribution. 

There is an uncertainty due to the Dalitz plot efficiency shape because of 
the difference in average efficiency over each bin for the flavor and \bdk samples. 
A maximum difference of 1.5\% is obtained in a MC study. 
The uncertainty is taken to be the maximum of the two comparable in size quantities: 
\begin{itemize}
  \item the root mean square (RMS) of $x$ and $y$ from smearing the numbers of 
events in the flavor sample $K_i$ by 1.5\% (larger for $y$ parameters); 
  \item the bias of $x$ and $y$ between fits with and without 
efficiency correction for $K_i$ obtained from signal MC (larger for $x$ parameters). 
\end{itemize}

The uncertainty due to cross-feed of events between bins is 
estimated by varying the momentum resolution by 20\% --- the MC 
resolution scaling factor obtained from the fit to \bdpi\ sample ---
and by taking the bias between the fits with and without final state 
interactions taken into account. 

The uncertainty arising from the finite sample of flavor-tagged 
$\dkpp$ decays is evaluated by varying the values of $K_i$ 
within their statistical errors. 

The final results for $(x,y)$ are corrected for the fit bias obtained from 
fits of MC pseudo-experiments. The uncertainty due to the fit bias 
is taken from the difference of biases for various input values of $x$ and $y$. 

The uncertainty due to the limited precision of the $c_i$ and $s_i$ parameters 
is obtained by smearing the $c_i$ and $s_i$ values within their 
total errors and repeating the fits for the same experimental data. 
We have performed a study of this procedure using both MC pseudo-experiments and 
analytical calculations. We find that the uncertainty obtained in 
this way is sample-dependent for small $B$ data samples and its average value 
scales in inverse proportion to the square root of the sample size. 
It reaches a constant value for large $B$ data samples 
(in the systematics-dominated case). This explains the somewhat higher 
uncertainty compared to the CLEO estimate given in~\cite{cleo_2}, which was obtained 
in the limit of a very large $B$ sample. 
In addition, the uncertainty in $(x,y)$ is proportional to $r_B$, and, thus,  
the uncertainty in the phases $\phi_3$ and $\delta_B$ is independent of 
$r_B$. As a result, the uncertainty of $(x,y)$ in the \bdk sample fit is 
3--4 times larger than in the \bdpi sample. 

\begin{table*}
  \caption{Systematic uncertainties in the $(x,y)$ measurements for \bdpi and \bdk samples
           in units of $10^{-3}$.}
  \label{tab:syst}
  \begin{tabular}{|l|cccc|cccc|}
    \hline
                                         & \multicolumn{4}{c|}{\bdpi} & \multicolumn{4}{c|}{\bdk} \\

    \cline{2-9}

    Source of uncertainty                & $\Delta x_-$ & $\Delta y_-$ & $\Delta x_+$ & $\Delta y_+$
                                         & $\Delta x_-$ & $\Delta y_-$ & $\Delta x_+$ & $\Delta y_+$ \\
    \hline

    Signal shape                         & $0.9$ & $1.9$  & $1.1$  & $5.0$
                                         & $7.3$ & $7.4$  & $7.3$  & $5.1$ \\
 
    $u,d,s,c$ continuum background       & $0.9$ & $1.4$  & $0.8$  & $1.3$
                                         & $6.7$ & $5.6$  & $6.6$  & $3.2$ \\
 
    $B\overline{B}$ background           & $3.3$ & $1.6$  & $4.5$  & $1.1$
                                         & $7.7$ & $8.4$  & $7.4$  & $5.4$ \\

    \bdpi background                     & $-$   & $-$    & $-$    & $-$
                                         & $1.2$ & $4.2$  & $1.9$  & $1.9$ \\

    Dalitz plot efficiency               & $3.0$ & $1.9$  & $3.2$  & $1.6$
                                         & $4.8$ & $2.0$  & $5.6$  & $2.1$ \\

    Crossfeed between bins               & $0.3$ & $0.6$  & $0.1$  & $0.7$
                                         & $0.0$ & $3.9$  & $0.1$  & $1.0$ \\

    Flavor-tagged statistics             & $1.7$ & $2.0$  & $1.6$  & $2.0$
                                         & $1.5$ & $2.7$  & $1.7$  & $1.9$ \\

    Fit bias                             & $0.4$ & $0.5$  & $0.4$  & $0.5$
                                         & $3.2$ & $5.8$  & $3.2$  & $5.8$ \\

    $c_i$ and $s_i$ precision            & $2.6$ & $6.5$  & $2.6$  & $6.5$
                                         & $10.1$& $22.5$ & $7.2$  & $17.4$\\

    \hline

    Total without $c_i$,$s_i$ precision  & $ 4.9$& $4.1$  & $6.0$  &$5.9$
                                         & $14.0$& $15.3$ & $14.1$ &$10.6$ \\

    \hline

    Total                                & $5.6$ & $7.7$  & $ 6.5$ & $8.8$
                                         & $17.3$& $27.2$ & $15.9$ & $20.4$ \\

    \hline
  \end{tabular}
\end{table*}

\section{Results for $\phi_3$, $r_B$ and $\delta_B$}

We use the frequentist approach with the Feldman-Cousins ordering~\cite{fc} to 
obtain the physical parameters $\mu = (\phi_3, r_B, \delta_B)$ from the 
measured parameters $z=(x_-, y_-, x_+, y_+)$, as was done in 
previous Belle analyses~\cite{belle_phi3_2,belle_phi3_3}. 
In essence, the confidence level $\alpha$
for a set of physical parameters $\mu$ is calculated as 
\begin{equation}
  \alpha(\mu) = \int\limits_{\mathcal{D}(\mu)}p(z|\mu)dz\left/ 
                \int\limits_{\infty}p(z|\mu)dz\right., 
\end{equation}
where $p(z|\mu)$ is the probability density to obtain the measurement 
result $z$ given the set of physics parameters $\mu$. The integration 
domain $\mathcal{D}(\mu)$ is given by the likelihood ratio (Feldman-Cousins)
ordering: 
\begin{equation}
  \frac{p(z|\mu)}{p(z|\mu_{\rm best}(z))} > 
  \frac{p(z_0|\mu)}{p(z_0|\mu_{\rm best}(z_0))}, 
\end{equation}
where $\mu_{\rm best}(z)$ is $\mu$ that maximizes $p(z|\mu)$ 
for the given $z$, and $z_0$ is the result of the data fit. 

In contrast to previous Belle analyses~\cite{belle_phi3_2, belle_phi3_3}, the probability 
density $p(z|\mu)$ is a multivariate Gaussian PDF with the errors 
and correlations between $x_{\pm}$ and $y_{\pm}$ taken from the 
data fit result. In the previous analyses, this PDF was taken from 
MC pseudo-experiments. 

\begin{figure}
  \parbox[t]{0.5\textwidth}{
  \includegraphics[width=0.242\textwidth]{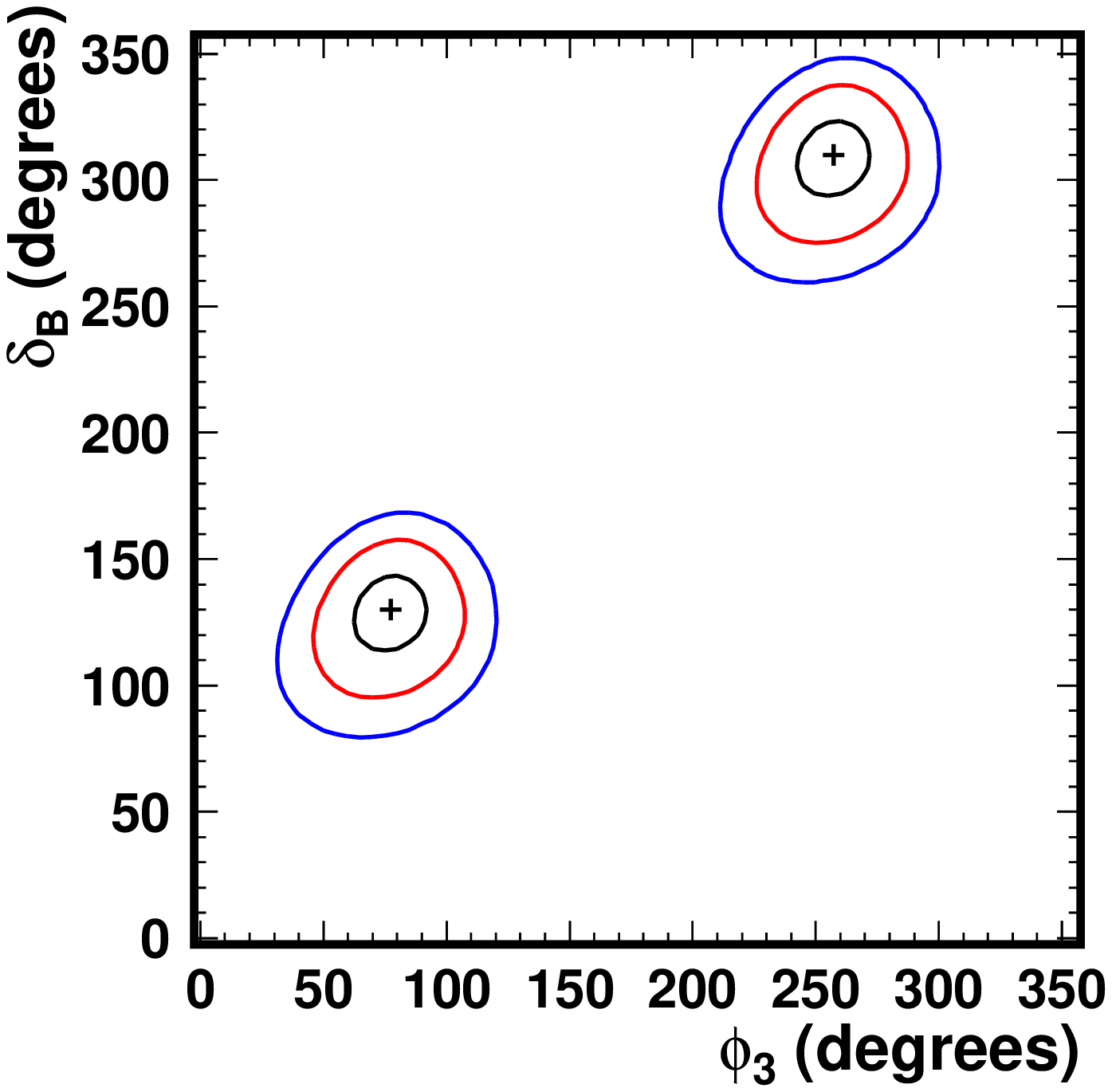}
  \includegraphics[width=0.242\textwidth]{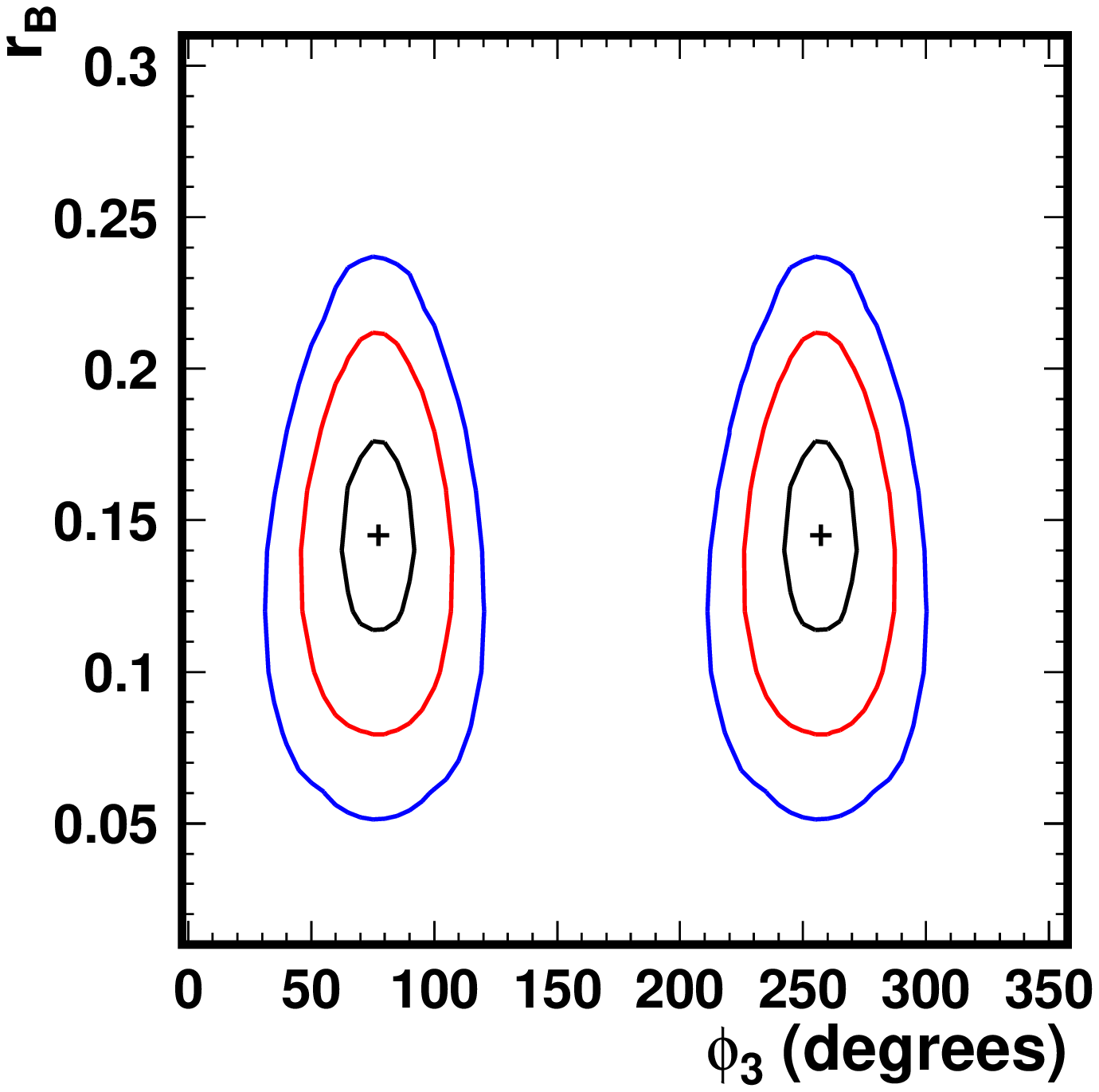}
  \put(-225,102){(a)}
  \put(-97,102){(b)}
  }
  \caption{Two-dimensional projections of 
           confidence region onto $(\phi_3, \delta_B)$ and 
           $(\phi_3, r_B)$ planes (one, two, and three standard 
           deviations, statistical only). }
  \label{fig:bdk_cl}
\end{figure}

As a result of this procedure, we obtain the confidence levels (CL) 
for the set of 
physical parameters $\phi_3, r_B$, and $ \delta_B$. The confidence levels 
for one and two standard deviations are taken at 20\% and 74\%
(appropriate for the case of a three-dimensional Gaussian distribution). The projections 
of the 3D surfaces bounding one, two, and three standard deviations 
volumes onto $(\phi_3, r_B)$ and $(\phi_3, \delta_B)$
planes are shown in Fig.~\ref{fig:bdk_cl}. 

Systematic errors in $\mu$ are obtained by varying the measured 
parameters $z$ within their systematic errors (Gaussian distributions are assumed) 
and calculating the RMS of $\mu_{\rm best}(z)$. In this calculation we 
assume that the systematic errors are uncorrelated between the $B^+$ and $B^-$ samples. 
In the case of 
$c_i$ and $s_i$ systematics, we test this assumption. When the fluctuation 
in $c_i$ and $s_i$ is generated, we perform fits to both $B^+$ and 
$B^-$ data with the same fluctuated $(c_i, s_i)$. We observe no 
significant correlation between the resulting $x_-$ and $x_+$ ($y_-$ and $y_+$). 

The final results are:
\begin{equation}
\begin{split}
\phi_3 & = (77.3^{+15.1}_{-14.9} \pm 4.1 \pm 4.3)^{\circ}\\
r_B & = 0.145\pm 0.030 \pm 0.010\pm 0.011\\
\delta_B & = (129.9\pm 15.0\pm 3.8\pm 4.7)^{\circ}, 
\end{split}
\end{equation}
where the first error is statistical, the second is systematic 
error without $c_i$ and $s_i$ uncertainty, 
and the third error is due to $c_i$ and $s_i$ uncertainty. 
Extraction of $\phi_3, r_B$, and $ \delta_B$ has a two-fold 
ambiguity, $(\phi_3,\delta)$ and $(\phi_3+180\deg, \delta+180\deg)$, 
leading to the same values of $x_{\pm}$ and $y_{\pm}$. Here we choose 
the solution that satisfies $0<\phi_3<180^{\circ}$.

The significance of $CP$ violation ($\phi_3$ being non-zero) is 
calculated as the CL of the point $\phi_3=0$. This calculation accounts for
a small deviation from Gaussian errors for $x$ and $y$ observed and 
parameterized using a large number of MC pseudo-experiments. The statistical 
significance equals $99.64$\% or 2.9 standard deviations. This value is 
in good agreement with the $\chi^2$ probability from the 
difference of the number of events in bins for $B^+$ and $B^-$ data. 
With the systematic uncertainties included, the significance decreases 
to $99.35$\% or 2.7 standard deviations. 

\section{Conclusion}

We report the results of a measurement of the Unitarity Triangle angle 
$\phi_3$ using a model-independent Dalitz plot analysis of
\dkpp\ decay in the process \bdk. 
The measurement was performed 
with the full data sample of 711 fb$^{-1}$ 
($772\times 10^6$ $B\overline{B}$ pairs) collected by the Belle detector
at the $\Upsilon(4S)$ resonance. 
Model independence is achieved by binning the Dalitz plot of the \dkpp 
decay and using the strong-phase coefficients for bins measured by 
the CLEO experiment~\cite{cleo_2}. 
We obtain the value
$\phi_3 = (77.3^{+15.1}_{-14.9} \pm 4.1 \pm 4.3)^{\circ}$;
of the two possible solutions we choose the one with $0<\phi_3<180^{\circ}$.
We also obtain the value of the amplitude ratio 
$r_B = 0.145\pm 0.030 \pm 0.010\pm 0.011$. 
In both results, the first error is statistical, the second is systematic 
error without $c_i$ and $s_i$ uncertainty, 
and the third error is due to $c_i$ and $s_i$ uncertainty.

This analysis is the first application of a novel method for measuring $\phi_3$. 
Compared to the result of the model-dependent measurement performed by Belle
with the \bdk mode, $\phi_3=(80.8^{+13.1}_{-14.8} \pm 5.0(\mbox{syst})
\pm 8.9(\mbox{model}))^{\circ}$~\cite{belle_phi3_3}, 
this measurement has somewhat poorer statistical precision despite a larger data sample 
used. There are two factors responsible for lower statistical sensitivity: 
1) the statistical error for the same statistics is inversely proportional 
to the $r_B$ value, and the central value of $r_B$ in this 
analysis is smaller, and 2) the binned approach is expected to have the statistical 
precision that is, on average, 10--20\% poorer than the unbinned one~\cite{modind2008}.

More important is that the large model uncertainty of the model-dependent 
result ($8.9^{\circ}$) is replaced by the 
purely statistical uncertainty of $4.3^{\circ}$ due to the limited size of the 
CLEO $\psi(3770)$ data sample.
Although the model-independent approach does not offer significant improvement 
over the unbinned model-dependent Dalitz plot analysis with the current data sample, 
it is promising for future measurements 
at super flavor factories~\cite{superkekb, superb} and LHCb~\cite{lhcb_modind}.
We expect that the statistical error of the $\phi_3$ measurement 
using the statistics of a 50 ab$^{-1}$ data sample that will be available at a 
super-B factory 
will reach 1--2$^{\circ}$. With the use of BES-III data~\cite{besiii}
the error due to the phase terms in the \dkpp decay will decrease to 
$1^{\circ}$ or less. We also expect that the experimental systematic 
error can be kept at the level below $1^{\circ}$, since most of 
its sources are limited by the statistics of the control channels. 

\section*{Acknowledgments}

We thank the KEKB group for the excellent operation of the
accelerator; the KEK cryogenics group for the efficient
operation of the solenoid; and the KEK computer group,
the National Institute of Informatics, and the 
PNNL/EMSL computing group for valuable computing
and SINET4 network support.  We acknowledge support from
the Ministry of Education, Culture, Sports, Science, and
Technology (MEXT) of Japan, the Japan Society for the 
Promotion of Science (JSPS), and the Tau-Lepton Physics 
Research Center of Nagoya University; 
the Australian Research Council and the Australian 
Department of Industry, Innovation, Science and Research;
the National Natural Science Foundation of China under
contract No.~10575109, 10775142, 10875115 and 10825524; 
the Ministry of Education, Youth and Sports of the Czech 
Republic under contract No.~LA10033 and MSM0021620859;
the Department of Science and Technology of India; 
the Istituto Nazionale di Fisica Nucleare of Italy; 
the BK21 and WCU program of the Ministry Education Science and
Technology, National Research Foundation of Korea,
and GSDC of the Korea Institute of Science and Technology Information;
the Polish Ministry of Science and Higher Education;
the Ministry of Education and Science of the Russian
Federation and the Russian Federal Agency for Atomic Energy;
the Slovenian Research Agency;  the Swiss
National Science Foundation; the National Science Council
and the Ministry of Education of Taiwan; and the U.S.\
Department of Energy and the National Science Foundation.
This work is supported by a Grant-in-Aid from MEXT for 
Science Research in a Priority Area (``New Development of 
Flavor Physics''), and from JSPS for Creative Scientific 
Research (``Evolution of Tau-lepton Physics'').

This research is partially funded by the Russian Presidential Grant 
for support of young scientists, grant number MK-1403.2011.2. 

\section*{Appendix}

\begin{figure*}
  \includegraphics[width=\textwidth]{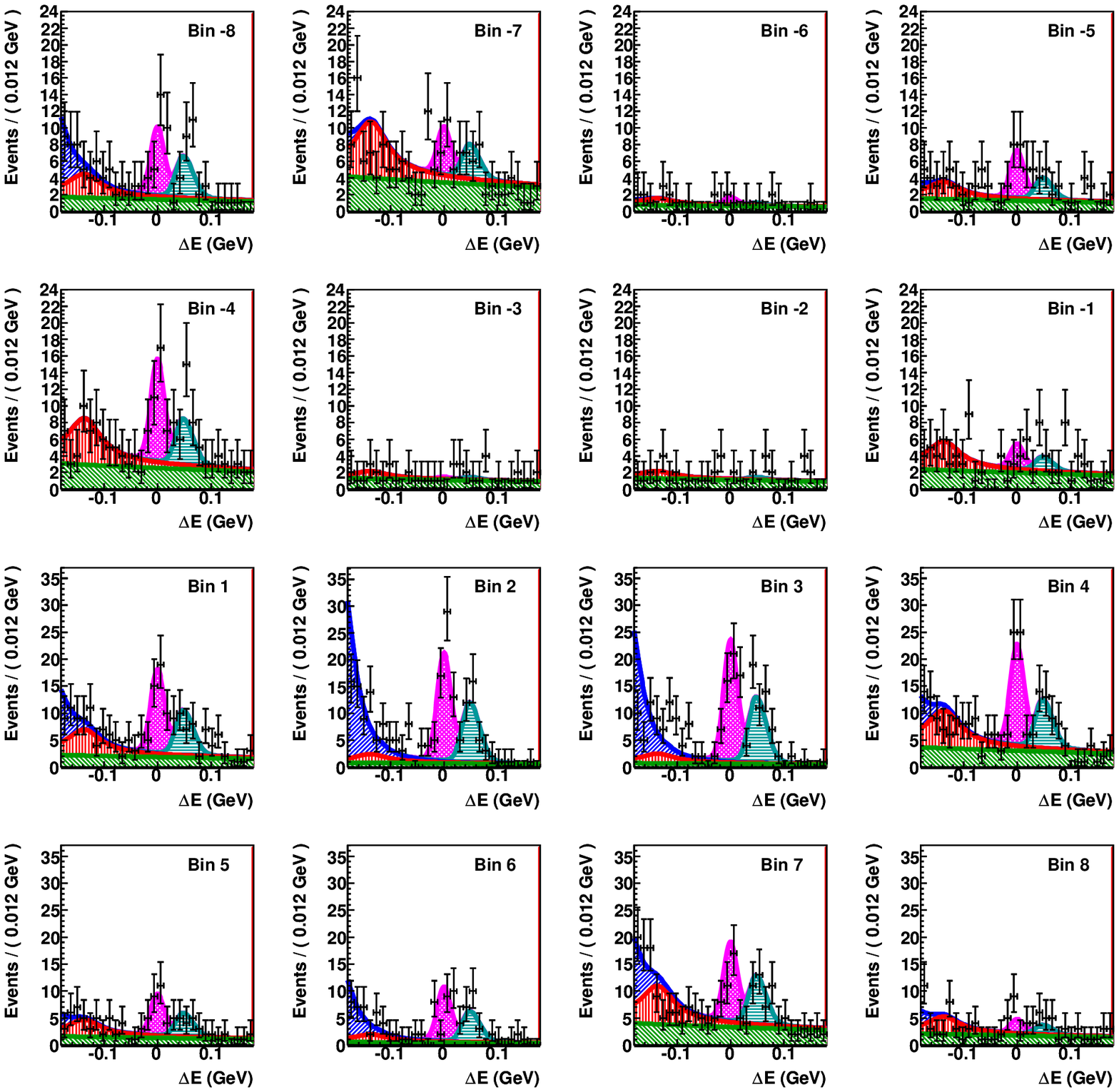}
  \caption{Projections of the combined fit of the \bdkp sample on $\de$
  for each Dailtz plot bin, 
  with the $\mbc>5270$~MeV/$c^2$ and $\thr<0.8$ requirements. 
  The fill styles for the signal and background components are the same 
  as in Fig.~\ref{fig:bdk_exp_all}. }
  \label{fig:fit_p}
\end{figure*}

\begin{figure*}
  \includegraphics[width=\textwidth]{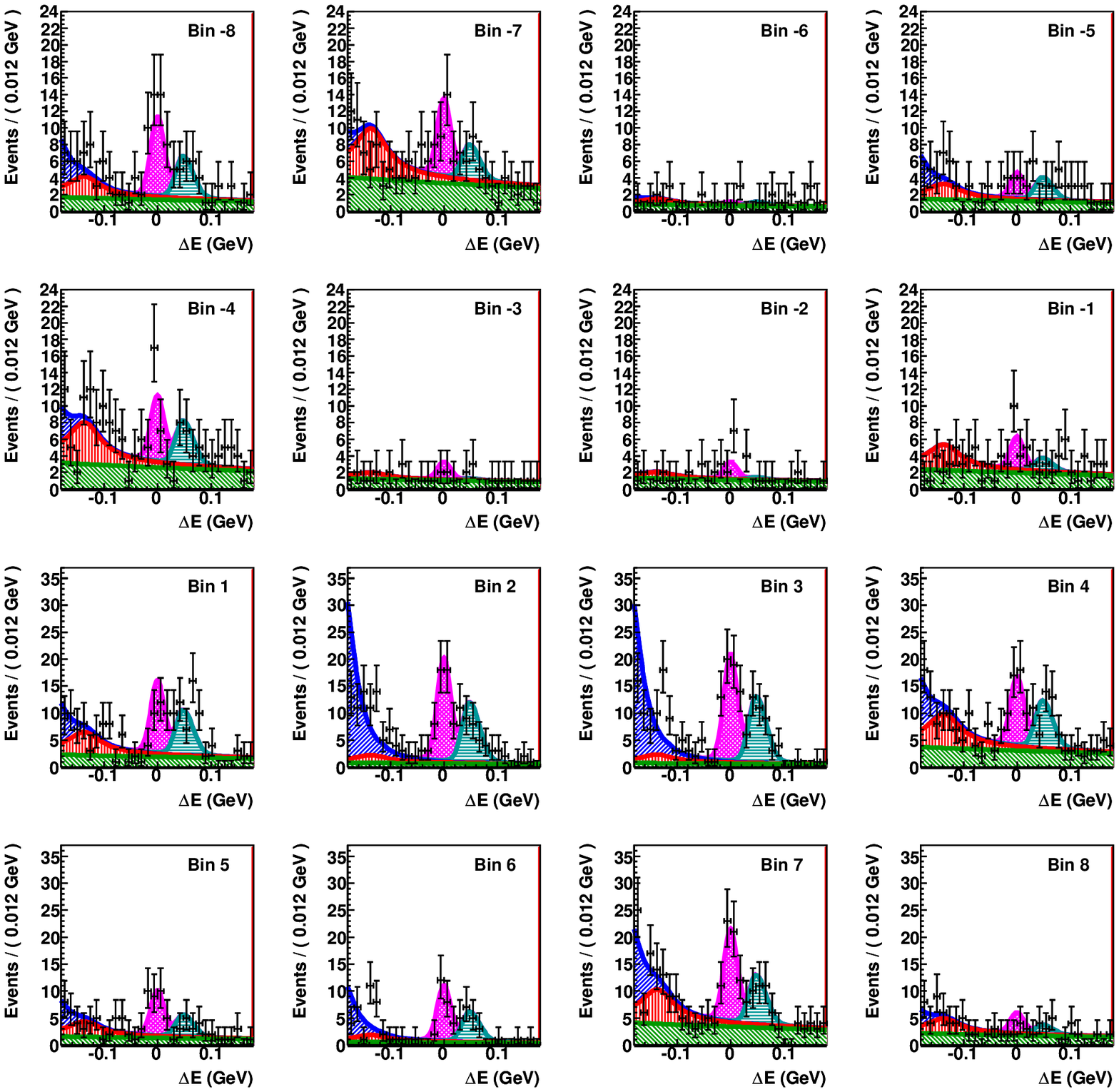}
  \caption{Projections of the combined fit of the \bdkm sample on $\de$
  for each Dailtz plot bin, 
  with the $\mbc>5270$~MeV/$c^2$ and $\thr<0.8$ requirements. 
  The fill styles for the signal and background components are the same 
  as in Fig.~\ref{fig:bdk_exp_all}.}
  \label{fig:fit_m}
\end{figure*}

The results of the combined fit to \bdkp and \bdkm samples separately 
for each bin of the Dalitz plot are shown in Figs.~\ref{fig:fit_p} and \ref{fig:fit_m}, 
respectively. The plots show the projections of the data and the fitting model 
on the \de variable, with the additional requirements $\mbc>5270$~MeV/$c^2$ and $\thr<0.8$. 

\end{document}